\documentclass[amssymb,aps,prd,twocolumn,nofootinbib,superscriptaddress]{revtex4-2}
\usepackage{graphicx}
\usepackage{subcaption} 
\usepackage{mathtools}
\usepackage{slashed,bm}
\usepackage{url}
\usepackage[normalem]{ulem}
\usepackage{booktabs} 
\usepackage{enumitem}
\usepackage{mathrsfs}
\usepackage{float}
\usepackage{ragged2e}
\usepackage[colorlinks=true, linkcolor=blue, citecolor=red, urlcolor=magenta]{hyperref}
\usepackage{amsmath,amssymb,bm}
\usepackage{hyperref}
 \usepackage{pgfplotstable}
\usepackage{tikz,pgfplots}
\usetikzlibrary{pgfplots.groupplots} 
\pgfplotsset{compat=1.18}
\usepgfplotslibrary{fillbetween}
\usepgfplotslibrary{groupplots}
\usepackage{comment}
\usepackage{xcolor}

\definecolor{ihqcdblue}{RGB}{68,119,170}
\definecolor{latticeorange}{RGB}{238,119,51}
\definecolor{refgray}{RGB}{120,120,120}
\definecolor{ihqcdblue}{RGB}{68,119,170}
\definecolor{softpurple}{RGB}{136,86,167}
\definecolor{latticeorange}{RGB}{238,119,51}
\definecolor{softred}{RGB}{204,102,119}
\definecolor{refgray}{RGB}{120,120,120}

\newcommand{\orcid}[1]{\href{https://orcid.org/#1}{\includegraphics[width=8pt]
		{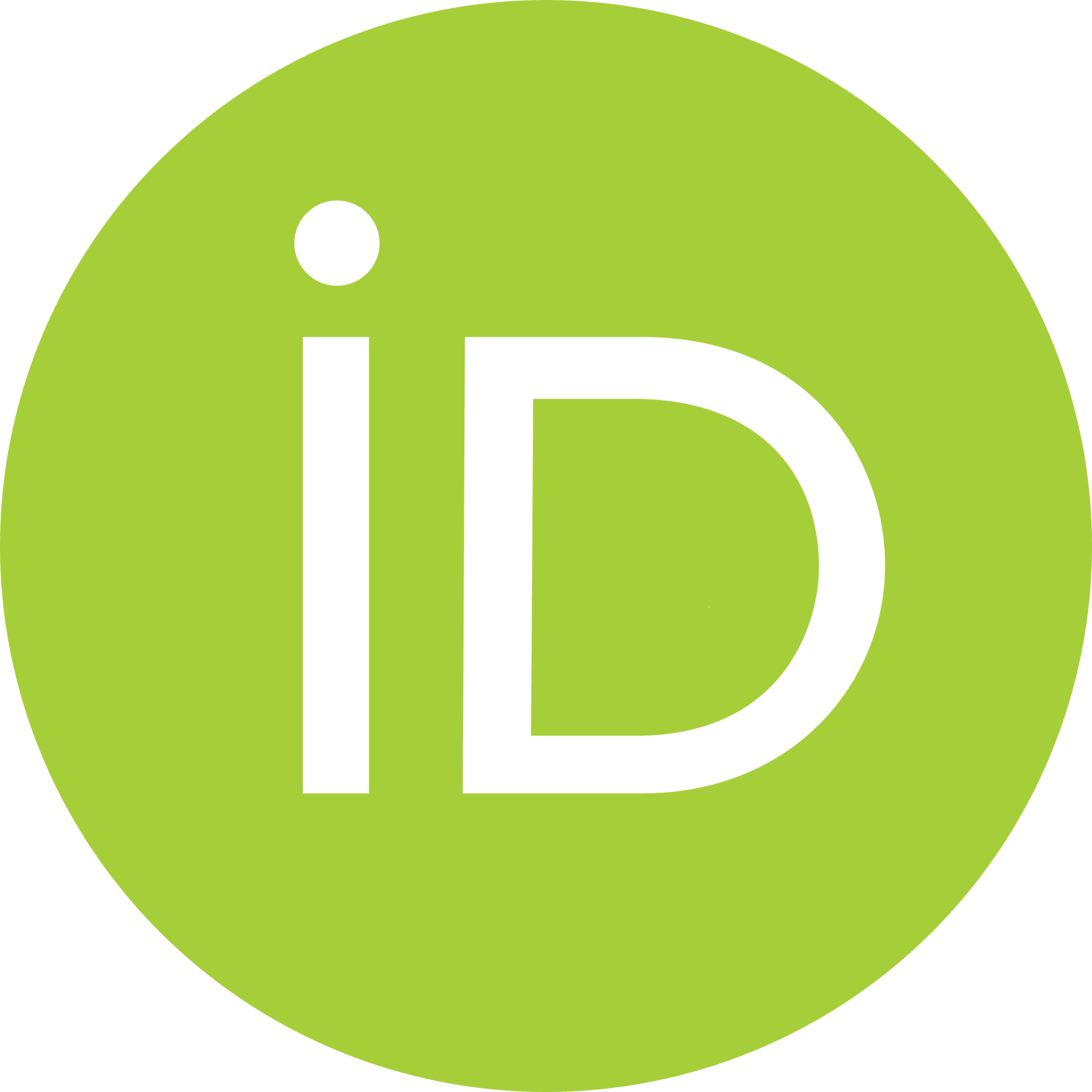}}}

\newcommand{\dd}{\mathop{}\!\mathrm{d}}

\pgfplotsset{
PRDaxis/.style={
    axis line style={black!70, line width=0.7pt},
    tick style={black!70, line width=0.6pt},
    tick label style={font=\small},
    label style={font=\small},
    title style={font=\small},
    legend style={
        font=\scriptsize,
        draw=none,
        fill=none,
        cells={anchor=west}
    },
    major grid style={black!8},
    grid=major,
    minor tick num=1,
    clip=false
}
}

\begin{document}

\title{ Minimal-Length Corrections to Hyperon Polarization in Rotating Quark Gluon Plasma}
\author{Behnam Pourhassan}
\affiliation{School of Physics, Damghan University, 3671641167 Damghan, Iran}
\affiliation{Center for Theoretical Physics, Khazar University, 41 Mehseti Street, Baku, AZ1096, Azerbaijan}
\author{Sameer Ahmad Mir\orcid{0000-0001-7528-1861}}
\email{sameerphst@gmail.com}
 \affiliation{Canadian Quantum Research Center, 460 Doyle Ave 106, Kelowna, BC V1Y 0C2, Canada} 
\affiliation{Department of Computer Sciences, Asian School of Business, Noida, Uttar Pradesh, 201303, India}
	
\begin{abstract}
Hyperon polarization in relativistic heavy-ion collisions provides a
sensitive probe of thermal vorticity and of the operator structure of
local equilibrium. We investigate an isotropic generalized uncertainty
principle (GUP) introduced through a spin-independent scalar deformation
of the one-particle phase-space density of states in the local rest frame
(LRF). The analysis is formulated with the Belinfante-Rosenfeld
local-equilibrium operator, the standard on-shell dispersion relation,
and the Pauli-Lubanski/axial-Wigner representation of the spin-$1/2$
observable, retaining terms linear in the GUP coefficient and in
hydrodynamic gradients. Within this framework, the same scalar GUP factor
appears in the scalar and axial phase-space contributions and cancels
from the local mean-spin ratio at fixed spacetime point and momentum.
Consequently, the leading GUP contribution to an experimentally relevant
polarization observable arises through the Cooper-Frye freeze-out
integrals and can be written in terms of a covariance between the LRF
momentum invariant and the local spin kernel. This cancellation is a
consequence of the scalar density-of-states modification within a fixed
pseudogauge and does not imply pseudogauge independence of the
polarization observable. For $\Lambda$ hyperons, a Maxwell-Boltzmann
thermal estimate shows that the dependence on freeze-out temperature and
fluid-dynamical correlations is substantially larger than the
Fermi-Dirac correction, which remains negligible over the temperature
range considered. Present LHC precision corresponds only to a formal
linear sensitivity of order $10$-$30~\mathrm{GeV}^{-2}$. Since this
interval lies beyond the domain in which the linear GUP expansion is
reliable, no exclusion limit is inferred. Differential measurements of
hyperon polarization as functions of momentum, rapidity, collision
centrality, and azimuthal harmonic structure provide the most direct
experimental avenue for testing the predicted covariance dependence.
\end{abstract}

\maketitle

\section{Introduction}

Non-central ultrarelativistic nucleus-nucleus collisions create quantum chromodynamics (QCD) matter with enormous angular momentum~\cite{STAR:2017ckg,Liang:2004ph}. A substantial fraction of this angular momentum is converted into rotational structures of the expanding quark-gluon plasma (QGP), making relativistic heavy-ion collisions a unique laboratory for studying vortical dynamics in strongly interacting matter~\cite{Pang:2016igs,Fu:2021pok,STAR:2023eck}. A decisive step in this direction was the experimental discovery of global $\Lambda$ and $\bar{\Lambda}$ hyperon polarization in Au+Au collisions at Relativistic Heavy-Ion Collider (RHIC), demonstrating that the QGP created in the laboratory is among the most vortical fluids ever observed \cite{STAR:2017ckg,STAR:2021beb}. Subsequent measurements from the RHIC Beam Energy Scan (BES) program have further resolved the collision energy and centrality dependence of global hyperon polarization, with the high-statistics BES-II data providing significantly improved experimental precision for confronting theoretical calculations of vortical spin polarization \cite{STAR:2023nvo}. 
At Large Hadron Collider (LHC) energies, measurements of global hyperon polarization remain compatible with zero within the present experimental uncertainties \cite{ALICE:2019onw}, while measurements of the polarization along the beam direction reveal nonvanishing local structures associated with anisotropic expansion and hydrodynamic gradients \cite{STAR:2019erd,ALICE:2021pzu,STAR:2023eck}. These observations demonstrate that hyperon polarization is sensitive not only to the overall vortical motion of the system but also to the detailed space-time structure of the QGP evolution \cite{ALICE:2019onw,ALICE:2021pzu,Becattini:2017gcx,Becattini:2021suc,Becattini:2021iol,Fu:2021pok}. These developments have established polarization as an increasingly
differential probe of QGP dynamics and have stimulated the development
of relativistic hydrodynamics with spin-hydrodynamic descriptions of relativistic nuclear collisions \cite{Becattini:2020ngo,Florkowski:2017ruc,
Weickgenannt:2020aaf,Hongo:2021ona,Becattini:2024uha,Huang:2024ffg,
Singh:2024cub}.

A consistent relativistic description of spin polarization in heavy-ion
collisions follows from quantum statistical mechanics at local thermodynamic
equilibrium. The local-equilibrium state is specified at the operator level
through a Zubarev-type density operator, while the leading spin response
is governed by gradients of the hydrodynamic fields. Of particular
importance is the thermal-vorticity tensor, which characterizes the
antisymmetric gradients of the inverse-temperature four-vector.
In the standard leading-order local-equilibrium formulation, the
mean spin vector of emitted spin-$1/2$ hadrons contains a contribution
linear in the thermal vorticity evaluated on the freeze-out hypersurface,
thereby relating the vortical motion of the QGP to experimentally
accessible hadron polarization
\cite{Becattini:2012tc,Becattini:2013fla,Becattini:2014yxa,Fang:2016vpj,Becattini:2016gvu,
Karpenko:2016jyx,Becattini:2017gcx,Wei:2018zfb}.
More generally, local-equilibrium polarization can also receive
first-gradient contributions associated with thermal shear and other
hydrodynamic gradients \cite{Becattini:2021suc,Becattini:2021iol,Liu:2021uhn,Yi:2021ryh}.  The development of relativistic spin hydrodynamics has further shown that a consistent theoretical description of polarization requires a precise specification of the local-equilibrium operator, the hydrodynamic frame, and the pseudo-gauge convention employed to separate orbital and spin angular momentum \cite{Becattini:2018duy,Speranza:2020ilk,Hattori:2019lfp,Fukushima:2020ucl,Hongo:2021ona,Gallegos:2021bzp,Daher:2022xon}.
In the present analysis, we adopt the Belinfante-Rosenfeld (BR) local-equilibrium operator
\cite{Becattini:2018duy,Liu:2021nyg},
retain the standard relativistic on-shell dispersion relation
\cite{Gao:2019znl,Weickgenannt:2019dks},
and describe the spin-$1/2$ observable through the Pauli-Lubanski vector and the axial component of the covariant Wigner function
\cite{Fang:2016vpj,Becattini:2020sww,Gao:2019znl,Weickgenannt:2019dks}.
The calculation is restricted to terms linear in the GUP coefficient and to leading order in hydrodynamic gradients
\cite{Fang:2016vpj,Gao:2019znl,Weickgenannt:2019dks,Liu:2021nyg}.
These choices define the relativistic spin sector used below and provide a controlled basis for isolating the modification associated with the GUP phase-space density of states
\cite{Kempf:1994su,Vakili:2012qt}.

{
In parallel, quantum-gravity phenomenology has developed effective
descriptions in which short-distance physics introduces a minimal length
\cite{Das:2008kaa,Maggiore:1993rv,Garay:1994en,Kempf:1994su,Hossenfelder:2012jw}.
A widely used example is the generalized uncertainty principle (GUP),
whose deformed phase-space algebra produces a momentum-dependent Jacobian
in the density of states \cite{Kempf:1994su,Hossenfelder:2012jw,Vakili:2012qt,Ali:2009zq,Chang:2001bm}. Such Jacobians and their consequences for
partition functions, equations of state, and scalar kinetic moments are
well established and are not new results of this work \cite{Chang:2001bm,Vakili:2012qt,Fityo:2008zz,Abbasiyan-Motlaq:2014kqa,Abbasiyan-Motlaq:2015gia,Ali:2014dfa,Vagenas:2019wzd,AbouSalem:2015wha}. In the present
application, the Jacobian modifies state counting in a local
grand-canonical ensemble: temperature \(T\) and chemical potentials \(\mu\) constrain
the mean energy and conserved charges, while the scalar weight modifies
the momentum measure \cite{Chang:2001bm,Benczik:2002tt,Kempf:1994su,Fityo:2008zz,Vakili:2012qt,Vagenas:2019wzd}.
The new question addressed here is how a spin-blind phase-space factor
appears in an operator-defined Zubarev local-equilibrium state \cite{Li:2025pef}, how it acts on the scalar and axial components of the spin-$1/2$ Wigner function \cite{Fang:2016vpj,Gao:2019znl,Weickgenannt:2019dks,Becattini:2020sww}, and
which part survives the Cooper-Frye ratio used to form an observed
polarization. We obtain three closely related results. First, the scalar GUP factor
cancels from the local mean-spin vector at fixed spacetime point and
momentum under the assumptions specified above. Second, after integration
over the Cooper-Frye freeze-out hypersurface, the leading GUP contribution
to the observable polarization can be expressed through a covariance
between the LRF momentum invariant and the local spin kernel. Third, we
derive an acceptance-dependent estimate of the corresponding
phenomenological sensitivity. The principal new result of this work
therefore lies in establishing, within the Zubarev local-equilibrium and
Wigner-function formalism, how a scalar GUP modification of phase-space
state counting enters hyperon polarization observables. The underlying
GUP modification of the phase-space density of states is already known;
the new physics addressed here is its consequence for the local spin
structure and for polarization after Cooper-Frye freeze-out 
\cite{Kempf:1994su,Chang:2001bm,Fityo:2008zz,Vakili:2012qt}. 
The local cancellation holds only under the simultaneous assumptions that
$W=W(\Pi^2)\mathbf 1_{\mathrm{spin}}$ is isotropic in the LRF and acts
identically on all spin states, that the standard mass-shell condition
$p^2=m^2$ and the one-particle spin operator are retained \cite{Fang:2016vpj,Gao:2019znl,Weickgenannt:2019dks,Becattini:2020sww}, and that the
Belinfante-Rosenfeld pseudo-gauge is kept fixed \cite{Becattini:2018duy,Speranza:2020ilk,
Buzzegoli:2021wlg}. We further restrict the
analysis to linear order in $\beta_G$ and to leading order in hydrodynamic
gradients \cite{Fang:2016vpj,Liu:2021nyg,Li:2025pef}. Since the local spin vector is itself of first order in gradients,
the leading GUP contribution to the polarization observable is of order
$\beta_G\partial$. The cancellation need not persist if the GUP factor
acquires momentum-space anisotropy or spin dependence, if the dispersion
relation or spin operator is modified, if another pseudo-gauge is adopted \cite{Becattini:2018duy,Speranza:2020ilk,Buzzegoli:2021wlg},
or if terms beyond linear order in $\beta_G$ or beyond leading order in
gradients are retained \cite{Li:2020eon}. Under the assumptions stated above, the scalar GUP
factor cancels from the local mean-spin vector at fixed spacetime point and
momentum, while a nonvanishing contribution can remain after Cooper-Frye
freeze-out through the covariance of $\Pi^2$ with the local spin kernel over
the freeze-out hypersurface.
\\
The paper is organized as follows. Sec.~\ref{sec2} introduces the
isotropic GUP and the associated modification of the phase-space density
of states. Sec.~\ref{sec3} summarizes the local grand-canonical
relations required for the subsequent analysis. Sec.~\ref{sec4}
specifies the local-equilibrium density operator and develops the expansion
to leading order in hydrodynamic gradients. Sec.~\ref{sec5} derives the
local cancellation of the scalar GUP factor within the fixed
Belinfante-Rosenfeld pseudo-gauge, while Sec.~\ref{sec:freezeout} obtains
the momentum-differential and acceptance-averaged covariance expressions
after Cooper-Frye freeze-out. Sec.~\ref{sec:bounds} examines the
phenomenological sensitivity and the range over which the linear GUP
expansion remains applicable. Sec.~\ref{sec8} discusses the LRF
formulation, pseudo-gauge dependence, and the assumptions underlying the
analysis. Finally, Sec.~\ref{sec9} summarizes the main results.
}

\section{Choice of GUP Model and Deformed Phase Space}
\label{sec2}
A generalized uncertainty principle (GUP) incorporates a minimal length into the quantum commutation relations, leading to corresponding modifications of the phase-space structure, statistical mechanics, and relativistic kinetic theory~\cite{Kempf:1994su,Fityo:2008zz,Vakili:2012qt}. In the present work, the GUP deformation is treated perturbatively, keeping only terms linear in the small parameter(s) that suppress short-distance physics. Since the phenomenology of thermal vorticity and spin polarization is formulated in a LRF of each fluid element, we implement the GUP as a deformation of the spatial phase space in that LRF, and later assemble fully covariant expressions by writing all observable corrections as scalar weights multiplying standard covariant tensors \cite{Becattini:2013fla,Becattini:2014yxa}. This is the standard and technically consistent way used when minimal-length deformations are implemented into relativistic thermodynamics \cite{Kempf:1994su,Chang:2001bm,Benczik:2002tt,Hossenfelder:2012jw}.

A widely used multi-dimensional GUP that preserves rotational invariance is the Kempf-Mangano-Mann (KMM) algebra \cite{Kempf:1994su,Chang:2001bm}
\begin{align}
[x^i,p^j]
&=
i\left[
(1+\beta_G p^2)\delta^{ij}
+\beta'_G p^i p^j
\right].
\label{eq:GUP_general}\\
[p_i,p_j]&=0,
\quad
p^2\equiv \bm{p}\cdot \bm{p}.
\label{eq:pp0}
\end{align}
{To avoid confusion with the inverse-temperature four-vector
$\beta^\mu=u^\mu/T$, we denote the dimensionful GUP coefficients by
$\beta_G$ and $\beta_G'$. Restoring $\hbar$ and $c$, their conventional
Planck normalization is given by
\begin{equation}
\beta_G=\frac{\beta_0}{(M_{\mathrm{Pl}}c)^2}
       =\frac{\beta_0\ell_{\mathrm{Pl}}^2}{\hbar^2},
\quad
\beta_G'=\frac{\beta_0'}{(M_{\mathrm{Pl}}c)^2}
        =\frac{\beta_0'\ell_{\mathrm{Pl}}^2}{\hbar^2},
\end{equation}
where $M_{\mathrm{Pl}}=\sqrt{\hbar c/G}$,
$\ell_{\mathrm{Pl}}=\sqrt{\hbar G/c^3}$, and $\beta_0,\beta_0'$ are
dimensionless. In natural units, $\hbar=c=1$, this becomes
$\beta_G=\beta_0/M_{\mathrm{Pl}}^2$ and
$\beta_G'=\beta_0'/M_{\mathrm{Pl}}^2$. This normalization is conventional i.e.,
the heavy-ion calculation treats $\beta_G$ as an effective low-energy
coefficient and does not assume $\beta_0=\mathcal O(1)$ or identify the
effective scale with $M_{\mathrm{Pl}}$.}

A key point is that in $d\ge 2$ spatial dimensions, the Jacobi identities generally induce a non-vanishing $[x_i,x_j]$ unless $\beta'_G$ is related to $\beta_G$ \cite{Kempf:1994su,Chang:2001bm,Benczik:2002tt}. To leading order in the deformation, one finds the schematic structure as
\begin{equation}
[x_i,x_j]
=
i\,(2\beta_G-\beta'_G)\,(p_i x_j-p_j x_i)+\mathcal{O}(\beta^2_G,\beta_G\beta'_G,\beta'^2_G),
\label{eq:xx_structure}
\end{equation}
so that the choice
\begin{equation}
\beta'_G=2\beta_G
\label{eq:betaPrimeChoice}
\end{equation}
ensures commuting coordinates up to $\mathcal{O}(\beta_G)$, i.e.\ $[x_i,x_j]=\mathcal{O}(\beta^2_G)$, which is sufficient for all computations consistently truncated at first order in $\beta_G$ \cite{Kempf:1994su,Chang:2001bm}. In what follows we therefore adopt the {Jacobi-consistent isotropic} specialization given in Eq. \eqref{eq:betaPrimeChoice} and work to $\mathcal{O}(\beta_G)$. 
{
For an isotropic state with $\langle p_i\rangle=0$ and
$\Delta p_1=\Delta p_2=\Delta p_3\equiv\Delta p_i$, one has
$\langle\bm p^2\rangle=3(\Delta p_i)^2$. The diagonal Robertson inequality
therefore gives
\begin{equation}
\Delta x_i\Delta p_i\geq
\frac{\hbar}{2}
\left[1+(3\beta_G+\beta_G')(\Delta p_i)^2\right].
\end{equation}
Minimization with respect to $\Delta p_i$ yields
\begin{equation}
(\Delta x_i)_{\min}
=\hbar\sqrt{3\beta_G+\beta_G'}
=\sqrt{3\beta_0+\beta_0'}\,\ell_{\mathrm{Pl}}
=\sqrt{5\beta_0}\,\ell_{\mathrm{Pl}}.
\end{equation}
The three-dimensional isotropy relation is essential in obtaining this
componentwise minimum.
}
For statistical mechanics and kinetic theory, the relevant object is the deformed phase-space volume element implied by the GUP. In the classical limit, commutators reduce to the corresponding Poisson-bracket, \(\frac{1}{i}[\,\cdot\,,\,\cdot\,]\;\longrightarrow\;\{\,\cdot\,,\,\cdot\,\},\)
so that Eq. \eqref{eq:GUP_general} becomes
\begin{equation}
\{x_i,p_j\}
=
\big(1+\beta_G p^2\big)\delta_{ij}+\beta'_G p_ip_j,
\quad
\{p_i,p_j\}=0.
\label{eq:PB}
\end{equation}
Let $A_{ij}(p)\equiv \{x_i,p_j\}$, and the Liouville theorem in a deformed symplectic manifold implies that the invariant phase-space measure is modified by the determinant of $A_{ij}$ \cite{Chang:2001bm,Benczik:2002tt}
\begin{equation}
\dd \Gamma_{\rm GUP}
=
\frac{\dd^3x\,\dd^3p}{(2\pi)^3}\,
\frac{1}{\det A(p)}.
\label{eq:Liouville_measure_general}
\end{equation}
Since $A_{ij}$ consists of a scalar matrix plus a rank-one contribution, its determinant can be evaluated analytically as
\begin{equation}
A_{ij}=(1+\beta_G p^2)\delta_{ij}+\beta'_G p_ip_j,
\end{equation}
one finds in $3$ spatial dimensions
\begin{equation}
\det A
=
(1+\beta_G p^2)^2\Big(1+(\beta_G+\beta'_G)p^2\Big).
\label{eq:detA}
\end{equation}
Therefore the GUP-deformed density of states becomes
\begin{equation}
\dd \Gamma_{\rm GUP}
=
\frac{\dd^3x\,\dd^3p}{(2\pi)^3}\,
\frac{1}{(1+\beta_G p^2)^2\big(1+(\beta_G+\beta'_G)p^2\big)}.
\label{eq:measure_general}
\end{equation}
Specializing to the Jacobi-consistent isotropic choice $\beta'_G=2\beta_G$ yields
\begin{equation}
\dd \Gamma_{\rm GUP}
=
\frac{\dd^3x\,\dd^3p}{(2\pi)^3}\,
\frac{1}{(1+\beta_G p^2)^2(1+3\beta_G p^2)}.
\label{eq:measure_beta2beta}
\end{equation}
Expanding to first order gives the explicit $\mathcal{O}(\beta_G)$ suppression of high-momentum states,
\begin{equation}
\dd \Gamma_{\rm GUP}
\simeq
\frac{\dd^3x\,\dd^3p}{(2\pi)^3}\left(1-5\beta_G p^2\right).
\label{eq:measure_expanded}
\end{equation}
The dimensionless control parameter is $\beta_G p^2$, hence the expansion is justified for $p^2\ll \beta_G^{-1}$.  
If one {additionally} imposes $\beta'_G=0$ as an effective isotropic truncation, then Eq.  \eqref{eq:detA} reduces to $\det A=(1+\beta_G p^2)^3$ and the frequently quoted measure
\begin{equation}
\dd \Gamma_{\rm GUP}
=
\frac{\dd^3x\,\dd^3p}{(2\pi)^3}\,
\frac{1}{(1+\beta_G p^2)^3}
\simeq \frac{\dd^3x\,\dd^3p}{(2\pi)^3}\left(1-3\beta_G p^2\right)
\label{eq:measure_truncation}
\end{equation}
is recovered \cite{Chang:2001bm}. In the present work, however, we retain the Jacobi-consistent choice used in Eq. \eqref{eq:betaPrimeChoice} and therefore using Eqs.  \eqref{eq:measure_beta2beta}-\eqref{eq:measure_expanded} throughout. 
For a species of degeneracy $g$, the number of one-particle quantum states in $\dd^3x\,\dd^3p$ is
\begin{multline}
\dd N
=
g\,\dd \Gamma_{\rm GUP}
=
g\,\frac{\dd^3x\,\dd^3p}{(2\pi)^3}\,
\frac{1}{(1+\beta_G p^2)^2(1+3\beta_G p^2)}\\
\simeq
g\,\frac{\dd^3x\,\dd^3p}{(2\pi)^3}\left(1-5\beta_G p^2\right).
\label{eq:state_counting}
\end{multline}
This formula is the starting point for all subsequent GUP-modified thermodynamic and kinetic integrals in the rotating QGP, and it makes manifest that GUP reduces the available phase-space volume at large momenta, consistent with the minimal-length interpretation \cite{Kempf:1994su,Chang:2001bm,Hossenfelder:2012jw,Tawfik:2014zca}.

{
\section{GUP-modified local grand-canonical equilibrium}
\label{sec3}
In this section, the GUP contribution to the local grand-canonical equilibrium state is incorporated through the modified phase-space structure. The modification is expressed in terms of the Lorentz-scalar momentum variable $\Pi^2=(p\cdot u)^2-m^2$, which reduces to $\bm p^2$ in the LRF. The usual Fermi-Dirac and Bose-Einstein forms of the equilibrium distribution remain unchanged, while the GUP dependence appears in the momentum-space factor $\mathcal W(\Pi^2)$. 
The positive-energy Lorentz-invariant on-shell momentum measure used
throughout this work is defined as
\begin{equation}
dP \equiv
\frac{d^3\mathbf{p}}{(2\pi)^3 E_{\mathbf p}}
=
2\,\frac{d^4p}{(2\pi)^3}\,
\theta(p^0)\,\delta(p^2-m^2),
\label{eq.3.1}
\end{equation}
where $E_{\mathbf p}\equiv\sqrt{\mathbf{p}^2+m^2}$. 
The factor of two in the four-dimensional representation follows from
\begin{equation}
\delta(p^2-m^2)
=
\frac{1}{2E_{\mathbf p}}
\left[
\delta(p^0-E_{\mathbf p})
+
\delta(p^0+E_{\mathbf p})
\right].
\end{equation}
After imposing the positive-energy condition $p^0>0$, this gives
\begin{equation}
2\,\theta(p^0)\,\delta(p^2-m^2)
=
\frac{1}{E_{\mathbf p}}\,
\delta(p^0-E_{\mathbf p}),
\end{equation}
so that the four-dimensional and three-dimensional representations
of the on-shell measure in Eq.~\eqref{eq.3.1} are exactly equivalent. 
The momentum dependence of the GUP factor can be expressed covariantly in terms of
\begin{equation}
\Pi^2=(p\cdot u)^2-m^2.
\label{eq:Pi2_cov}
\end{equation}
In the LRF, $u^\mu=(1,\bm 0)$, Eq.~\eqref{eq:Pi2_cov}
reduces to \(\Pi^2=\bm p^2.\)
For the GUP realization considered here, the corresponding scalar
phase-space factor is
\begin{equation}
\mathcal W(\Pi^2)
=
1-\kappa\beta_G\Pi^2
+\mathcal O(\beta_G^2\Pi^4),
\quad
\kappa=5.
\label{eq:W_general}
\end{equation}
The value $\kappa=3$, associated with the frequently employed
$\beta_G'=0$ truncation, is not included in the numerical analysis.
It is quoted only to indicate that the coefficient of the linear term
depends on the particular GUP convention
\cite{Kempf:1994su,Chang:2001bm,Benczik:2002tt}.
Fig.~\ref{fig:Wfactor} shows the linear form of
Eq.~\eqref{eq:W_general} as a function of
$q=\sqrt{\Pi^2}/T_f$. The three curves correspond to
$\kappa\beta_GT_f^2=0$, $0.002$, and $0.02$, with
$0\leq q\leq2$. Within this interval, the largest correction obeys
$\kappa\beta_G\Pi^2\leq0.08$.
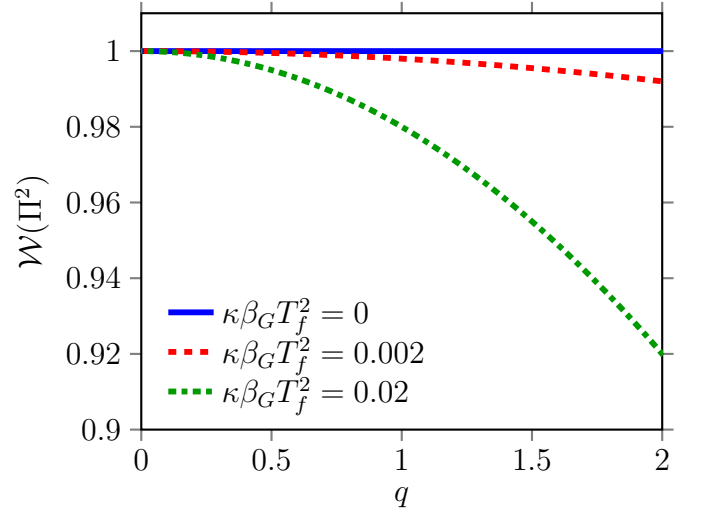
\begin{figure}[t]
\centering

\begin{tikzpicture}
\begin{axis}[
    width=0.98\columnwidth,
    height=0.82\columnwidth,
    xmin=0,
    xmax=2,
    ymin=0.90,
    ymax=1.01,
    xlabel={$q$},
    ylabel={$\mathcal W(\Pi^2)$},
    xtick={0,0.5,1.0,1.5,2.0},
    ytick={0.90,0.92,0.94,0.96,0.98,1.00},
    tick align=outside,
    axis line style={thick},
    tick style={thick},
    xlabel style={font=\large},
    ylabel style={font=\large},
    tick label style={font=\large},
    legend style={
        at={(0.03,0.03)},
        anchor=south west,
        draw=none,
        fill=none,
        font=\large
    },
    legend cell align={left},
    samples=200,
]

\addplot[
    blue,
    line width=2.2pt,
    solid
]
{1};
\addlegendentry{$\kappa\beta_G T_f^2=0$}

\addplot[
    red,
    line width=2.2pt,
    dashed
]
{1-0.002*x^2};
\addlegendentry{$\kappa\beta_G T_f^2=0.002$}

\addplot[
    green!60!black,
    line width=2.2pt,
    dashdotted
]
{1-0.02*x^2};
\addlegendentry{$\kappa\beta_G T_f^2=0.02$}

\end{axis}
\end{tikzpicture}

\caption{\justifying Scalar GUP factor
$\mathcal W(\Pi^2)=1-(\kappa\beta_GT_f^2)q^2$ as a function of
$q=\sqrt{\Pi^2}/T_f$ for
$\kappa\beta_GT_f^2=0$, $0.002$, and $0.02$.
The interval $0\leq q\leq2$ is displayed, for which the largest
deformation satisfies $\kappa\beta_G\Pi^2\leq0.08$.
The curves represent the spin-independent momentum dependence
introduced by the GUP factor. The linear approximation is restricted
to the region $\kappa|\beta_G|\Pi^2\ll1$.}
\label{fig:Wfactor}

\end{figure}
The local grand-canonical distribution is written covariantly as
\begin{equation}
f_\eta(x,p)
=
\frac{1}{
\exp\!\left[
\beta_\mu(x)p^\mu-\sum_a q_a\xi_a(x)
\right]+\eta},
\label{eq:f_covariant}
\end{equation}
where $\beta^\mu(x)={u^\mu(x)}/{T(x)}$ and $\xi_a(x)={\mu_a(x)}/{T(x)}$ where $q_a$ denotes the corresponding charge eigenvalue and
\begin{equation}
\eta=+1
\quad\text{for fermions},
\qquad
\eta=-1
\quad\text{for bosons}.
\label{eq:eta_def}
\end{equation}
Equivalently,
\begin{align}
f_+(x,p)
&=
\left[
e^{\beta\cdot p-\sum_a q_a\xi_a}+1
\right]^{-1}
&&\text{for fermions},
\label{eq:f_fermion}
\\
f_-(x,p)
&=
\left[
e^{\beta\cdot p-\sum_a q_a\xi_a}-1
\right]^{-1}
&&\text{for bosons}.
\label{eq:f_boson}
\end{align}
The factor $\mathcal W(\Pi^2)$ leaves the functional form of the
Fermi-Dirac and Bose-Einstein distributions in
Eqs.~\eqref{eq:f_fermion} and \eqref{eq:f_boson} unchanged.
Its contribution appears through the GUP-dependent phase-space factor in the momentum integrals \cite{Chang:2001bm,Benczik:2002tt,Tawfik:2014zca}.
For the local-equilibrium operator and the freeze-out calculation, the
relevant quantities are the particle four-current and the
energy-momentum tensor is given as
\begin{align}
N^\mu_{\rm GUP}(x)
&=
g\int \dd P\,
p^\mu\mathcal W(\Pi^2)f_\eta(x,p),
\label{eq:Nmu_GUP}
\\
T^{\mu\nu}_{\rm GUP}(x)
&=
g\int \dd P\,
p^\mu p^\nu\mathcal W(\Pi^2)f_\eta(x,p).
\label{eq:Tmunu_GUP}
\end{align}
Eqs.~\eqref{eq:Nmu_GUP} and \eqref{eq:Tmunu_GUP} provide the
quantities required for the local-equilibrium operator introduced in
the next section. Their covariant form is necessary because
$u^\mu(x)$, and consequently $\Pi^2(x,p)$, changes across the
freeze-out hypersurface. No further thermodynamic moments are required
for the polarization calculation developed below.
}

\section{Local Equilibrium Density Operator}
\label{sec4}

For a strongly interacting relativistic fluid, local equilibrium is most cleanly defined at the operator level, without assuming an underlying quasiparticle kinetic description. The appropriate framework is the Zubarev nonequilibrium statistical operator (NSO) method, where the (quasi-)local equilibrium density operator is obtained by maximizing the von Neumann entropy under constraints fixing the local expectation values of conserved currents on a spacelike hypersurface $\Sigma$ \cite{Becattini:2012tc,Becattini:2014yxa,Hayata2015PRD,Uddin:2011bi}. 
{ In the scalar GUP implementation adopted here, the microscopic local
operators remain unchanged. We therefore retain the standard
Belinfante-Rosenfeld stress-energy tensor
$\widehat T_{\rm BR}^{\mu\nu}$ and the conserved current
$\widehat j^\mu$ in the Zubarev local-equilibrium operator. The GUP
modification is introduced through the one-particle phase-space factor
\(dP_{\rm GUP}=dP\,\mathcal{W}(\Pi^2),\) which modifies the momentum integrals defining expectation values and kinetic moments. Accordingly, the quantities
$T_{\rm GUP}^{\mu\nu}(x)$ and $N_{\rm GUP}^{\mu}(x)$ introduced in
Sec.~\ref{sec3} represent macroscopic moments obtained with the
GUP-dependent phase-space factor, rather than new microscopic operators.
This distinction is essential because $\mathcal{W}(\Pi^2)$ depends on the local
hydrodynamic field $u^\mu(x)$ and therefore belongs to the local
statistical description of the fluid element, rather than to the
microscopic definition of the stress-energy tensor. The Zubarev
local-equilibrium operator thus retains its standard functional form,
while the GUP dependence appears in the expectation values and in the
relations connecting the local thermodynamic fields to the macroscopic
currents.
} 
Let $\hat{\rho}$ be a density operator. The von Neumann entropy is
\begin{equation}
S[\hat{\rho}] = - \mathrm{Tr}\!\left(\hat{\rho}\,\ln \hat{\rho}\right),
\label{eq:vonNeumannEntropy}
\end{equation}
to be maximized subject to normalization $\mathrm{Tr}\,\hat{\rho}=1$ and fixed mean densities of the conserved currents on $\Sigma$
\begin{align}
\mathrm{Tr}\!\left(
\widehat\rho\,\widehat T_{\rm BR}^{\mu\nu}(x)
\right)
&=
T^{\mu\nu}(x)
\label{eq:constraint_T}\\
\mathrm{Tr}\!\left(\hat{\rho}\,\hat{j}^{\mu}(x)\right) &= j^{\mu}(x).
\label{eq:constraint_j}
\end{align}
Here the quantities on the right-hand sides denote the local mean
densities fixed by the hydrodynamic fields, their kinetic expressions
are evaluated with the GUP-dependent phase-space factor introduced in
Sec.~\ref{sec3}. Introducing the Lagrange-multiplier fields
$\beta_\nu(x)$ and $\xi(x)$, the constrained functional takes the form
\begin{multline}
\Phi[\widehat\rho]
=
S[\widehat\rho]
-\lambda
\left(\mathrm{Tr}\widehat\rho-1\right)
-\int_\Sigma d\Sigma_\mu\times\\
\Bigg[
\beta_\nu\,
\mathrm{Tr}\!\left(
\widehat\rho\,\widehat T_{\rm BR}^{\mu\nu}
\right)
-
\xi\,
\mathrm{Tr}\!\left(
\widehat\rho\,\widehat j^\mu
\right)
\Bigg].
\label{eq:constrained_entropy_functional}
\end{multline}
The stationarity condition $\delta\Phi/\delta\hat{\rho}=0$ yields
\begin{equation}
\ln \hat{\rho}_{\rm LE}
=
-\lambda
-\int_\Sigma \dd\Sigma_\mu \left(
\hat{T}^{\mu\nu}_{\rm BR}\beta_\nu
-\xi \hat{j}^\mu
\right),
\end{equation}
hence
\begin{equation}
\widehat\rho_{\rm LE}
=
\frac{1}{Z_{\rm LE}}
\exp\!\left[
-\int_\Sigma d\Sigma_\mu
\left(
\widehat T_{\rm BR}^{\mu\nu}\beta_\nu
-
\xi\,\widehat j^\mu
\right)
\right].
\label{eq:rho_LE_def}
\end{equation}
The microscopic operators employed in the Zubarev formalism satisfy the standard local conservation laws,
\begin{equation}
\partial_\mu\widehat T_{\rm BR}^{\mu\nu}=0,
\quad
\partial_\mu\widehat j^\mu=0.
\label{eq:conservation_assumption}
\end{equation}
These conservation laws are the standard operator identities required
for the hypersurface formulation of the Zubarev formalism. The present
GUP ansatz leaves them unchanged. The GUP dependence is restricted to
the phase-space factor used in evaluating the corresponding local mean
quantities. This assumption is commonly adopted in effective GUP
thermodynamics, where the modification is introduced through the
phase-space structure while the underlying spacetime symmetries are
retained in the region considered.  
For generic space-dependent $\beta_\mu(x)$ and $\xi(x)$, $\hat{\rho}_{\rm LE}$ depends on the choice of $\Sigma$. Consider two hypersurfaces $\Sigma_1$ and $\Sigma_2$ bounding a four-volume $\Omega$. Using Gauss' theorem,
\begin{multline}
\int_{\Sigma_2}\!\dd\Sigma_\mu\left(\hat{T}^{\mu\nu}_{\rm BR}\beta_\nu-\xi\hat{j}^\mu\right)
-\int_{\Sigma_1}\!\dd\Sigma_\mu\left(\hat{T}^{\mu\nu}_{\rm BR}\beta_\nu-\xi\hat{j}^\mu\right)
=\\
\int_{\Omega}\!\dd^4x\,\partial_\mu\left(\hat{T}^{\mu\nu}_{\rm BR}\beta_\nu-\xi\hat{j}^\mu\right)
\\
=
\int_{\Omega}\!\dd^4x\left[
(\partial_\mu\beta_\nu)\hat{T}^{\mu\nu}_{\rm BR}
-(\partial_\mu\xi)\hat{j}^\mu
\right],
\label{eq:hypersurface_dependence}
\end{multline}
where Eq. \eqref{eq:conservation_assumption} was used. 
If the stress-energy tensor entering the exponent is taken to be the symmetric Belinfante tensor (assumption adopted here), then $\hat{T}^{\mu\nu}_{\rm BR}=\hat{T}^{\nu\mu}_{\rm BR}$ and only the symmetric part of $\partial_\mu\beta_\nu$ contributes
\begin{equation}
(\partial_\mu\beta_\nu)\hat{T}^{\mu\nu}_{\rm BR}
=
\frac{1}{2}\big(\partial_\mu\beta_\nu+\partial_\nu\beta_\mu\big)\hat{T}^{\mu\nu}_{\rm BR}.
\label{eq:Killing_contraction}
\end{equation}
Therefore, $\hat{\rho}_{\rm LE}$ becomes hypersurface-independent if
\begin{equation}
\partial_\mu\beta_\nu+\partial_\nu\beta_\mu=0,
\qquad
\partial_\mu\xi=0,
\label{eq:global_equilibrium_conditions}
\end{equation}
namely if $\beta^\mu$ is a Killing vector field and $\xi$ is constant. Under these conditions, the system is in global thermodynamic equilibrium (possibly with acceleration and rotation), and the density operator reduces to a generalized Gibbs ensemble with conserved generators \cite{Becattini:2012tc,Becattini:2014yxa}. 
A general solution of the Killing equation in Minkowski spacetime is
\begin{equation}
\beta_\mu(x)=b_\mu+\varpi_{\mu\nu}\,x^\nu,
\label{eq:beta_global_solution}
\end{equation}
where $b_\mu=\mathrm{const}$, $\varpi_{\mu\nu}=\mathrm{const}$, $\varpi_{\mu\nu}=-\varpi_{\nu\mu}$ and $\varpi_{\mu\nu}$ is constant thermal vorticity. Since the microscopic operator adopted here is the standard Belinfante-Rosenfeld stress-energy tensor, the corresponding conserved generators on $\Sigma$ are
\begin{equation}
\widehat P^\nu_{\rm BR}
\equiv
\int_{\Sigma} d\Sigma_\mu\,
\widehat T_{\rm BR}^{\mu\nu}(x),
\label{eq.4.12}
\end{equation}
\begin{equation}
\widehat J_{\rm BR}^{\rho\sigma}
\equiv
\int_{\Sigma} d\Sigma_\mu
\left[
x^\rho\widehat T_{\rm BR}^{\mu\sigma}(x)
-
x^\sigma\widehat T_{\rm BR}^{\mu\rho}(x)
\right],
\end{equation}
and
\begin{equation}
\widehat Q
\equiv
\int_{\Sigma} d\Sigma_\mu\,
\widehat j^\mu(x).
\label{eq.4.14}
\end{equation}
The global-equilibrium density operator therefore takes the form
\begin{equation}
\widehat\rho_{\rm GE}
=
\frac{1}{Z_{\rm GE}}
\exp\left[
-b_\mu\widehat P^\mu_{\rm BR}
+\frac{1}{2}\varpi_{\rho\sigma}
\widehat J_{\rm BR}^{\rho\sigma}
+\xi\widehat Q
\right].
\label{eq.4.15}
\end{equation}
The subscript ${\rm BR}$ specifies the local operator representation
used in the present formalism. When the boundary terms associated
with a pseudogauge transformation vanish, the corresponding integrated
energy-momentum and total angular-momentum charges agree with the
global conserved charges in other pseudogauges. 
It is convenient to decompose gradients of $\beta_\mu$ into antisymmetric and symmetric parts, away from the global equilibrium
\begin{align}
\varpi_{\mu\nu} &\equiv -\frac{1}{2}\big(\partial_\mu\beta_\nu-\partial_\nu\beta_\mu\big),
\label{eq:varpi_def}\\
\zeta_{\mu\nu} &\equiv \frac{1}{2}\big(\partial_\mu\beta_\nu+\partial_\nu\beta_\mu\big),
\label{eq:zeta_def}
\end{align}
where $\zeta_{\mu\nu}$ is the thermal shear tensor (symmetric). The basic identity is then
\begin{equation}
\partial_\mu\beta_\nu = -\varpi_{\mu\nu}+\zeta_{\mu\nu}.
\label{eq:partial_beta_decomp}
\end{equation}
Using $\beta^\mu=u^\mu/T$, one may rewrite thermal vorticity explicitly in terms of gradients of $u^\mu$ and $T$ as
\begin{equation}
\varpi_{\mu\nu}
=
-\frac{1}{2T}\big(\partial_\mu u_\nu-\partial_\nu u_\mu\big)
-\frac{1}{2T^2}\big(u_\mu \partial_\nu T-u_\nu \partial_\mu T\big),
\label{eq:varpi_uT}
\end{equation}
which makes clear that $\varpi_{\mu\nu}$ is dimensionless. 
Let $x$ be a reference point and writing the fields in the exponent of Eq. \eqref{eq:rho_LE_def} as a Taylor expansion around $x$
\begin{align}
\beta_\nu(y) &= \beta_\nu(x) + (y-x)^\lambda \partial_\lambda\beta_\nu(x) + \mathcal{O}(\partial^2),
\label{eq:beta_Taylor}\\
\xi(y) &= \xi(x) + (y-x)^\lambda \partial_\lambda\xi(x) + \mathcal{O}(\partial^2),
\label{eq:xi_Taylor}
\end{align}
where $y\in\Sigma$. Substituting Eqs. \eqref{eq:beta_Taylor}-\eqref{eq:xi_Taylor} into the exponent of Eq. \eqref{eq:rho_LE_def}, and using $\widehat P_{\rm BR}^{\nu}$ and $\widehat Q$ defined in Eqs.~\eqref{eq.4.12} and \eqref{eq.4.14}, respectively, we introduce the translated first-moment operators about the reference point $x$ as
\begin{equation}
\widehat M_{\rm BR}^{\lambda\nu}(x)
\equiv
\int_{\Sigma} d\Sigma_\mu\,
(y-x)^\lambda
\widehat T_{\rm BR}^{\mu\nu}(y),
\end{equation}
and
\begin{equation}
\widehat N^\lambda(x)
\equiv
\int_{\Sigma} d\Sigma_\mu\,
(y-x)^\lambda
\widehat j^\mu(y).
\end{equation}
Then, up to first order in gradients,
\begin{multline}
\int_\Sigma \dd\Sigma_\mu\,
\left(
\hat{T}^{\mu\nu}_{\rm BR}\beta_\nu
-\xi \hat{j}^\mu
\right)
=
\beta_\nu(x)\hat{P}^\nu_{\rm BR}
-\\ \xi(x)\hat{Q}
+ \partial_\lambda\beta_\nu(x)\,\hat{M}^{\lambda\nu}_{\rm BR}(x)
-\partial_\lambda\xi(x)\,\hat{N}^{\lambda}(x)
+\mathcal{O}(\partial^2).
\label{eq:exponent_expand_1}
\end{multline}
The tensor $\hat{M}^{\lambda\nu}$ admits antisymmetric and symmetric parts. Introducing the (about $x$) Belinfante-Rosenfeld angular-momentum operator as
\begin{multline}
\hat{J}^{\lambda\nu}_{\rm BR}(x)
\equiv
\hat{M}^{\lambda\nu}_{\rm BR}(x)-\hat{M}^{\nu\lambda}_{\rm BR}(x)
=\\
\int_\Sigma \dd\Sigma_\mu
\left[
(y-x)^\lambda \hat{T}^{\mu\nu}_{\rm BR}(y)-(y-x)^\nu \hat{T}^{\mu\lambda}_{\rm BR}(y)
\right],
\label{eq:J_about_x}
\end{multline}
and the symmetric quadrupole operator is given by
\begin{equation}
\hat{D}^{\lambda\nu}_{\rm BR}(x)
\equiv
\hat{M}^{\lambda\nu}_{\rm BR}(x)+\hat{M}^{\nu\lambda}_{\rm BR}(x).
\label{eq:D_def}
\end{equation}
Using $\partial_\lambda\beta_\nu = -\varpi_{\lambda\nu}+\zeta_{\lambda\nu}$, one finds the identity
{
\begin{equation}
\partial_\lambda\beta_\nu M^{\lambda\nu}
=-\frac{1}{2}\varpi_{\lambda\nu}J^{\lambda\nu}
+\frac{1}{2}\zeta_{\lambda\nu}D^{\lambda\nu}.
\label{eq:betaM_decomp}
\end{equation}
where the antisymmetric contraction projects onto $\hat{J}^{\lambda\nu}$ and the symmetric contraction projects onto $\hat{D}^{\lambda\nu}$.  
The minus sign in the antisymmetric contribution follows from
$\partial_\lambda\beta_\nu=-\varpi_{\lambda\nu}
+\zeta_{\lambda\nu}$. When included in the negative exponent of the local-equilibrium operator, this contribution yields the standard $+\tfrac12\varpi_{\lambda\nu}J^{\lambda\nu}$ term.
Substituting the preceding decomposition of
$\partial_\lambda\beta_\nu$ into the hypersurface expansion and retaining
terms through first order in hydrodynamic gradients gives the
local-equilibrium operator
\begin{multline}
\widehat{\rho}_{\mathrm{LE}}^{\mathrm{BR}}
=
\frac{1}{Z_{\mathrm{LE}}}
\exp\!\Bigg[
-\beta_\nu(x)\widehat{P}_{\mathrm{BR}}^\nu
+\sum_a \xi_a(x)\widehat{Q}_a
+\\\frac{1}{2}\varpi_{\lambda\nu}(x)
\widehat{J}_{\mathrm{BR}}^{\lambda\nu}(x)
-\frac{1}{2}\zeta_{\lambda\nu}(x)
\widehat{D}_{\mathrm{BR}}^{\lambda\nu}(x)
\\+\sum_a \partial_\lambda \xi_a(x)\,
\widehat{N}_a^\lambda(x)
+\mathcal{O}(\partial^2)
\Bigg].
\label{eq:rho_LE_gradient}
\end{multline}
Here $\widehat{\rho}_{\mathrm{LE}}^{\mathrm{BR}}$ denotes the local-equilibrium operator in the fixed Belinfante-Rosenfeld pseudogauge adopted throughout this work. 
Thus, within the scalar GUP framework adopted here, the label ``GUP'' refers to the phase-space factor applied to the momentum moments and Cooper-Frye freeze-out integrals, rather than to any modification of the microscopic operators appearing in $\widehat{\rho}_{\rm LE}^{\rm BR}$. The polarization analysis retains the nondissipative contribution linear in the thermal vorticity $\varpi_{\lambda\nu}$. The thermal-shear and chemical potential gradient terms are displayed to specify the complete first-order gradient structure of the local-equilibrium operator, but they are not included in the thermal-vorticity contribution considered below. Equation~\eqref{eq:rho_LE_gradient} defines the local-equilibrium operator used in the Wigner-function analysis of Sec.~\ref{sec5}.
}

\section{Mean Spin Vector with GUP Corrections}
\label{sec5}
The observed polarization of spin-$1/2$ hadrons emitted from a vortical QGP is a linear-response effect of the local-equilibrium density operator to thermal vorticity gradients. In the covariant formulation, the mean spin four-vector at a spacetime point $x$ and on-shell momentum $p$ is obtained from the spin-dependent part of the local-equilibrium Wigner function (or, equivalently, from the one-particle spin density matrix derived from $\hat{\rho}_{\rm LE}$). The standard leading-gradient result is linear in thermal vorticity and is uniquely fixed by Lorentz covariance, antisymmetry of $\varpi_{\mu\nu}$, and the orthogonality constraint $p_\mu S^\mu=0$ \cite{Becattini:2013fla,Becattini:2014yxa,Becattini:2012tc}. In this section, we first rederive the standard result with all gamma-matrix steps shown explicitly and then incorporate the GUP contribution consistently to $\mathcal{O}(\beta_G)$. We show that, for an isotropic GUP, no additional tensor structures arise at this order, the modification appears solely through a momentum-dependent scalar factor. We adopt metric signature $g^{\mu\nu}=\mathrm{diag}(1,-1,-1,-1)$ and Levi-Civita tensor $\varepsilon^{0123}=+1$. The Dirac spin generator is given by
\begin{equation}
\Sigma^{\mu\nu}\equiv \frac{i}{4}\,[\gamma^\mu,\gamma^\nu],
\label{eq:Sigma_def}
\end{equation}
and the covariant spin (Pauli-Lubanski) operator for a massive spin-$1/2$ particle of mass $m$ is
\begin{equation}
\hat{s}^\mu(p)\equiv -\frac{1}{2m}\,\varepsilon^{\mu\nu\rho\sigma}\,p_\sigma\,\Sigma_{\nu\rho},
\qquad
p_\mu \hat{s}^\mu(p)=0.
\label{eq:PauliLubanski_def}
\end{equation}
We also define the dual thermal vorticity tensor
\begin{equation}
\widetilde{\varpi}^{\mu\nu}\equiv \frac{1}{2}\varepsilon^{\mu\nu\rho\sigma}\varpi_{\rho\sigma},
\,\,
\varepsilon^{\mu\nu\rho\sigma}p_\nu\varpi_{\rho\sigma}=2\,\widetilde{\varpi}^{\mu\nu}p_\nu.
\label{eq:varpi_dual}
\end{equation}
A key trace identity used below is
\begin{equation}
\mathrm{Tr}_D\!\left[\gamma^5\gamma^\mu\,\Sigma^{\alpha\beta}\,(\slashed{p}+m)\right]
=2\,\varepsilon^{\mu\alpha\beta\nu}\,p_\nu,
\label{eq:trace_identity}
\end{equation}
where $\slashed{p}\equiv \gamma_\lambda p^\lambda$ and $\mathrm{Tr}_D$ denotes the trace over Dirac indices. 
With the metric convention $g^{\mu\nu}=\mathrm{diag}(1,-1,-1,-1)$,
$\epsilon^{0123}=+1$, and
$\gamma^5\equiv i\gamma^0\gamma^1\gamma^2\gamma^3$, the relevant
gamma-matrix trace is 
\(
\mathrm{Tr}_D\!\left(
\gamma^5\gamma^\mu\gamma^\nu\gamma^\rho\gamma^\sigma
\right)
=
-4i\,\epsilon^{\mu\nu\rho\sigma}
\) \cite{Becattini:2013fla}. 
Using $\Sigma^{\alpha\beta}=(i/4)
[\gamma^\alpha,\gamma^\beta]$, together with
$\mathrm{Tr}_D(\gamma^5\gamma^\mu\gamma^\alpha\gamma^\beta)=0$,
one obtains
\begin{multline}
\mathrm{Tr}_D\!\left[
\gamma^5\gamma^\mu\Sigma^{\alpha\beta}
(\slashed{p}+m)
\right]
=\\
\frac{i}{4}\,p_\nu
\Big[
\mathrm{Tr}_D(
\gamma^5\gamma^\mu\gamma^\alpha\gamma^\beta\gamma^\nu)
-
\mathrm{Tr}_D(
\gamma^5\gamma^\mu\gamma^\beta\gamma^\alpha\gamma^\nu)
\Big]
=\\
2\,\epsilon^{\mu\alpha\beta\nu}p_\nu,
\end{multline}
which establishes Eq.~\eqref{eq:trace_identity} with the conventions adopted here. 
The (spinor) Wigner function of a Dirac field at local equilibrium is given by
\begin{equation}
W_{\alpha\beta}(x,p)\equiv
\int \dd^4y\,e^{ip\cdot y}\,
\left\langle
\bar{\psi}_\beta\!\left(x-\frac{y}{2}\right)\,
\psi_\alpha\!\left(x+\frac{y}{2}\right)
\right\rangle_{\rm LE},
\label{eq:Wigner_def}
\end{equation}
with $\langle \cdots\rangle_{\rm LE}\equiv \mathrm{Tr}(\hat{\rho}_{\rm LE}\cdots)$. 
At leading order in gradients (first order in $\partial\beta$), the local-equilibrium Wigner function for a massive spin-$1/2$ species admits the standard expansion \cite{Becattini:2013fla,Becattini:2014yxa}
\begin{multline}
W_{\rm LE}(x,p)
=
2\pi\,\delta(p^2-m^2)\,
\Big[
(\slashed{p}+m)\,f(x,p)
-\\ \frac{1}{2}\,f(x,p)\big(1-f(x,p)\big)\,
\varpi_{\alpha\beta}(x)\,\Sigma^{\alpha\beta}\,(\slashed{p}+m)
\Big]
+\mathcal{O}(\partial^2),
\label{eq:Wigner_LTE}
\end{multline}
where $f(x,p)\equiv f_+(x,p)$ is the fermionic local-equilibrium distribution defined in Eq.~\eqref{eq:f_fermion}. 
Eq.~\eqref{eq:Wigner_LTE} is the precise operator-level implementation of linear response of the local-equilibrium state to thermal vorticity, and it is equivalent to the Kubo expansion of $\hat{\rho}_{\rm LE}$ derived in the previous section \cite{Becattini:2012tc,Becattini:2014yxa}. 
Decomposing $\mathcal{W}$ in the Clifford basis, the scalar and axial-vector components are
\begin{align}
F(x,p) &\equiv \mathrm{Tr}_D\!\left[\mathcal{W}(x,p)\right],
\label{eq:F_def}\\
A^\mu(x,p) &\equiv \mathrm{Tr}_D\!\left[\gamma^\mu\gamma^5\,\mathcal{W}(x,p)\right].
\label{eq:A_def}
\end{align}
Using \eqref{eq:Wigner_LTE}, the scalar component is given by
\begin{multline}
F(x,p)=2\pi\,\delta(p^2-m^2)\,\mathrm{Tr}_D\!\left[(\slashed{p}+m)\right]\,f(x,p)
=\\ 2\pi\,\delta(p^2-m^2)\,(4m)\,f(x,p).
\label{eq:F_eval}
\end{multline}
The axial component receives no contribution from the zeroth-order term, and at first order one finds, using \eqref{eq:trace_identity},
\begin{multline}
A^\mu(x,p)
=
-2\pi\,\delta(p^2-m^2)\,\frac{1}{2}\,f(1-f)\,
\varpi_{\alpha\beta}\,\times\\
\mathrm{Tr}_D\!\left[\gamma^\mu\gamma^5\,\Sigma^{\alpha\beta}\,(\slashed{p}+m)\right]
+\mathcal{O}(\partial^2)
\\
=
-2\pi\,\delta(p^2-m^2)\,f(1-f)\,
\varpi_{\alpha\beta}\,
\varepsilon^{\mu\alpha\beta\nu}\,p_\nu
+\mathcal{O}(\partial^2).
\label{eq:A_eval}
\end{multline} 
The mean spin four-vector in phase space is then defined as can be seen in the Refs. \cite{Becattini:2013fla,Becattini:2014yxa}
\begin{equation}
S^\mu(x,p)\equiv \frac{1}{2}\,\frac{A^\mu(x,p)}{F(x,p)}.
\label{eq:S_from_Wigner}
\end{equation}
Combining Eq. \eqref{eq:F_eval}, Eq. \eqref{eq:A_eval}, and Eq. \eqref{eq:S_from_Wigner} yields
\begin{multline}
S^\mu(x,p)
=
-\frac{1}{8m}\,\big(1-f(x,p)\big)\,
\varepsilon^{\mu\alpha\beta\nu}\,p_\nu\,\varpi_{\alpha\beta}(x)\\
=
-\frac{1}{4m}\,\big(1-f(x,p)\big)\,
\widetilde{\varpi}^{\mu\nu}(x)\,p_\nu,
\label{eq:standardS}
\end{multline}
which is the standard leading-gradient thermal-vorticity polarization formula. The orthogonality constraint is immediate
\(
p_\mu S^\mu(x,p)=0,
\)
because $p_\mu p_\nu$ is symmetric while $\widetilde{\varpi}^{\mu\nu}$ is antisymmetric. 
{The following result is conditional on
\begin{multline}
\mathcal A_{\mathrm{scope}}=
\left\{
\begin{gathered}
\mathcal{W}=\mathcal{W}(\Pi^2)\mathbf 1_{\mathrm{spin}}
\ \text{ is isotropic and scalar}\\ \text{in the LRF},\\
p^2=m^2\ \text{and the Pauli-Lubanski/Wigner spin} \\
\text{observable are undeformed},\\
\mathfrak p=\mathrm{BR}\ \text{is fixed},\quad
\text{terms are retained through }\\
\mathcal O(\partial)\ \text{and}\ \mathcal O(\beta_G\partial)
\end{gathered}
\right\}.
\label{eq.5.15}
\end{multline}
Within the scalar GUP framework adopted here, the modification is introduced exclusively through the on-shell phase-space factor, $dP\rightarrow dP\,\mathcal{W}(\Pi^2)$. The pointwise microscopic Wigner kernel itself remains unchanged. Accordingly, $F_{\rm GUP}$ and $A^\mu_{\rm GUP}$ denote the scalar and axial phase-space densities after inclusion of the common GUP factor, rather than new microscopic Wigner components. Because the same spin-independent scalar factor multiplies both quantities, the present GUP framework gives
\begin{equation}
\begin{aligned}
\mathcal F_{\mathrm{GUP}}(x,p)
=\mathcal{W}(\Pi^2)\mathcal F_0(x,p),
\\
\mathcal A_{\mathrm{GUP}}^\mu(x,p)
=\mathcal{W}(\Pi^2)\mathcal A_0^\mu(x,p)
\label{eq.5.16}
\end{aligned}
\end{equation}
to the retained order. Therefore,
\begin{align}
S_{\mathrm{BR,GUP}}^\mu(x,p)
&=\frac{1}{2}
\frac{\mathcal A_{\mathrm{GUP}}^\mu(x,p)}
{\mathcal F_{\mathrm{GUP}}(x,p)}
\nonumber\\
&=\frac{1}{2}
\frac{\mathcal{W}(\Pi^2)\mathcal A_0^\mu(x,p)}
{\mathcal{W}(\Pi^2)\mathcal F_0(x,p)}
\nonumber\\
&=S_{\mathrm{BR},0}^\mu(x,p)
+\mathcal O(\beta_G^2\partial,\partial^2,
\beta_G\partial^2).
\label{5.18}
\end{align}
Eq.~\eqref{5.18} is therefore a consequence of the scalar GUP framework within a fixed pseudogauge, rather than an independent dynamical statement about a modified microscopic Wigner function. It shows that a spin-independent scalar GUP factor does not introduce additional local tensor structures at first order. This result does not imply equivalence of spin observables defined in different pseudogauges. The cancellation need not persist for an anisotropic or spin-dependent GUP factor, a modified mass shell or spin operator, a finite spin potential, a change of pseudogauge, higher orders in $\beta_G$, or higher-order hydrodynamic gradients.}

\section{Freeze-Out and Polarization Observables}
\label{sec:freezeout}

Experimental polarization is obtained by combining the local-equilibrium spin polarization
at leading order in thermal vorticity and particlization on a freeze-out hypersurface
via the Cooper-Frye formula \cite{Becattini:2013fla,Becattini:2016gvu}. In the present framework, the GUP modifies the density of states through a spin-independent scalar factor multiplying the invariant phase-space measure \cite{Kempf:1994su}.  

Let $\Sigma$ denote a three-dimensional space-like hypersurface, parameterized by
$\xi^a$ ($a=1,2,3$), with embedding $x^\mu=x^\mu(\xi^1,\xi^2,\xi^3)$. The directed
surface element can be written as
\begin{equation}
d\Sigma_\lambda
=
\epsilon_{\lambda\alpha\beta\gamma}\,
\frac{\partial x^\alpha}{\partial \xi^1}\,
\frac{\partial x^\beta}{\partial \xi^2}\,
\frac{\partial x^\gamma}{\partial \xi^3}\,d^3\xi,
\label{eq:dSigma_param}
\end{equation}
so that $p^\lambda d\Sigma_\lambda$ is a Lorentz scalar. 
{ Using the positive-energy invariant on-shell measure defined in
Eq.~\eqref{eq.3.1}, its representation in transverse-momentum and rapidity
variables is
\begin{equation}
\frac{d^3\bm p}{(2\pi)^3E_{\bm p}}
=\frac{p_T\,dp_T\,d\phi\,dy}{(2\pi)^3},
\label{eq:dP_cyl}
\end{equation}}
where $E_{\bm p}=m_T\cosh y,\, m_T=\sqrt{m^2+p_T^2}$ is the energy and transverse mass of a particle having mass $m$ and $y$ is rapidity and $(p_T,\phi)$ are transverse polar coordinates. 
The local-equilibrium distribution and the scalar GUP factor used below
are those defined in Eqs.~\eqref{eq:f_covariant} and \eqref{eq:W_general}, respectively. Since
$\Pi^2$ depends on the local flow field $u^\mu(x)$, the factor
$W(\Pi^2)$ is generally $(x,p)$ dependent over the freeze-out
hypersurface \cite{Kempf:1994su}. 
The GUP-modified differential yield is given by the Cooper-Frye integral \cite{Mir:2026ruk,Mir:2026cbr,Mir:2026ryy,Mir:2026rgu}
\begin{equation}
E_p\frac{dN_{\rm GUP}}{d^3p}
=
\frac{g}{(2\pi)^3}
\int_\Sigma d\Sigma_\lambda\,p^\lambda
f(x,p)\,W(\Pi^2).
\label{eq:yield_spec_expanded}
\end{equation}
For the spin-$1/2$ species considered here, we use the leading-gradient
thermal-vorticity mean-spin vector derived in Eq.~\eqref{eq:standardS}, with thermal
vorticity defined in Eq.~\eqref{eq:varpi_def}. 
The corresponding emitted spin spectrum becomes
\begin{multline}
E_p\frac{dS^\mu_{\rm GUP}}{d^3p}
=
\frac{g}{(2\pi)^3}
\int_\Sigma d\Sigma_\lambda\,p^\lambda
f(x,p)\,\times\\W(\Pi^2)\,S^\mu(x,p)
+
\mathcal O(\beta_G^2,\partial^2),
\label{eq:spin_spec_expanded}
\end{multline}
where the fixed-pseudogauge local equality established in
Eq.~\eqref{5.18} has been used. The momentum-differential polarization four-vector is then defined as
\begin{equation}
P^\mu_{\rm GUP}(p)\equiv
2\,
\frac{dS^\mu_{\rm GUP}/d^3p}{dN_{\rm GUP}/d^3p}.
\label{eq:Pmu_def_expanded}
\end{equation}
For fixed $p$, the freeze-out weight is given by
\begin{equation}
\omega(x,p)
\equiv
\frac{g}{(2\pi)^3}\,
d\Sigma_\lambda\,p^\lambda f(x,p),
\label{eq:omega_def}
\end{equation}
and defining the standard (undeformed) hypersurface average at fixed $p$ by
\begin{equation}
\langle X\rangle_{\Sigma|p}
\equiv
\frac{\int_\Sigma \omega(x,p)\,X(x,p)}{\int_\Sigma \omega(x,p)}.
\label{eq:avg_Sigma_p}
\end{equation}
Then Eqs.~\eqref{eq:yield_spec_expanded}-\eqref{eq:spin_spec_expanded} imply
\begin{equation}
P^\mu_{\rm GUP}(p)
=
2\,
\frac{\int_\Sigma \omega(x,p)\,\mathcal{W}(\Pi^2)\,S^\mu(x,p)}
{\int_\Sigma \omega(x,p)\,\mathcal{W}(\Pi^2)}
\equiv
2\,\langle S^\mu\rangle^{\rm (GUP)}_{\Sigma|p},
\label{eq:Pmu_as_weighted_avg}
\end{equation}
where $\langle\cdots\rangle^{\rm (GUP)}_{\Sigma|p}$ denotes an average including the common GUP factor $\omega \mathcal{W}$. 
Expanding $\mathcal{W}(\Pi^2)=1-\kappa\beta_G\Pi^2+\mathcal{O}(\beta^2_G)$ yields, to first order in $\beta_G$,
\begin{equation}
\begin{aligned}
P^\mu_{\rm GUP}(p)
&=
2\,
\frac{\langle S^\mu(1-\kappa\beta_G\Pi^2)\rangle_{\Sigma|p}}
{\langle (1-\kappa\beta_G\Pi^2)\rangle_{\Sigma|p}}
+\mathcal{O}(\beta^2_G\partial,\partial^2,\beta^2_G\partial^2)
\\
&=
2\,
\Big(\langle S^\mu\rangle_{\Sigma|p}
-\kappa\beta_G\Big[\langle \Pi^2S^\mu\rangle_{\Sigma|p}
-\\& \hspace{1.1cm}\langle \Pi^2\rangle_{\Sigma|p}\,\langle S^\mu\rangle_{\Sigma|p}\Big]\Big)
+\mathcal{O}(\beta^2_G\partial,\partial^2,\beta^2_G\partial^2)
\\
&=
P^\mu_{\rm std}(p)
-2\kappa\beta_G\,
{\rm Cov}_{\Sigma|p}\!\left(\Pi^2,S^\mu\right)
+\\& \hspace{4.3cm}\mathcal{O}(\beta^2_G\partial,\partial^2,\beta^2_G\partial^2),
\label{eq:Pmu_cov_Sigma_p}
\end{aligned}
\end{equation}
where $P^\mu_{\rm std}(p)\equiv 2\langle S^\mu\rangle_{\Sigma|p}$ and
\begin{equation}
{\rm Cov}_{\Sigma|p}\!\left(\Pi^2,S^\mu\right)
\equiv
\langle \Pi^2S^\mu\rangle_{\Sigma|p}
-\langle \Pi^2\rangle_{\Sigma|p}\,\langle S^\mu\rangle_{\Sigma|p}.
\label{eq:cov_Sigma_p_def}
\end{equation}
Therefore, within the isotropic scalar GUP framework, the leading $\mathcal{O}(\beta_G)$ contribution at fixed $p$ is governed by correlations over the freeze-out hypersurface arising from the spatial variation of $u^\mu(x)$, $T(x)$, and $\varpi_{\mu\nu}(x)$ across $\Sigma$. 
A useful special limit follows immediately, if $\Pi^2(x,p)$ is effectively constant over the dominant emission region on $\Sigma$ for that momentum bin, then
${\rm Cov}_{\Sigma|p}(\Pi^2,S^\mu)=0$, and Eq.~\eqref{eq:Pmu_cov_Sigma_p} reduces to
\begin{equation}
P^\mu_{\rm GUP}(p)=P^\mu_{\rm std}(p)+\mathcal{O}(\beta^2_G,\partial^2).
\label{eq:Pmu_diff_cancel_conditional}
\end{equation}
For later bounds, the correction can be bounded by Cauchy-Schwarz
\begin{multline}
\left|\Delta P^\mu(p)\right|
\equiv \left|P^\mu_{\rm GUP}(p)-P^\mu_{\rm std}(p)\right|
\le
2\kappa|\beta_G|\,\times\\
\sqrt{{\rm Var}_{\Sigma|p}(\Pi^2)}\,
\sqrt{{\rm Var}_{\Sigma|p}(S^\mu)}
+\mathcal{O}(\beta^2_G,\partial^2),
\label{eq:CS_bound_fixedp}
\end{multline}
with ${\rm Var}(X)\equiv \langle X^2\rangle-\langle X\rangle^2$. Let ${\cal A}$ denote an experimental acceptance region in momentum space, for instance
\begin{equation}
{\cal A}:\quad
y\in[y_{\min},y_{\max}],
\,\,
p_T\in[p_{T,\min},p_{T,\max}],
\,\,
\phi\in[0,2\pi).
\label{eq:acceptance_example}
\end{equation}
The polarization averaged over an experimental momentum-space acceptance
$\mathcal A$ is defined from the momentum-integrated spin and particle
spectra as
\begin{equation}
P^\mu_{\rm GUP}[{\cal A}]
\equiv
2\,\frac{\int_{\cal A} d^3p\,\frac{dS^\mu_{\rm GUP}}{d^3p}}
{\int_{\cal A} d^3p\,\frac{dN_{\rm GUP}}{d^3p}}.
\label{eq:Pmu_accept_expanded}
\end{equation}
{
Using Eqs.~\eqref{eq:yield_spec_expanded} and \eqref{eq:spin_spec_expanded}, the spectra appearing in Eq.~\eqref{eq:Pmu_accept_expanded} satisfy
\begin{equation}
\begin{aligned}
\frac{dN_{\rm GUP}}{d^3p}
=&
\frac{1}{E_{\mathbf p}}
\int_\Sigma
\omega(x,p)\,W(\Pi^2),
\\
\frac{dS^\mu_{\rm GUP}}{d^3p}
&=
\frac{1}{E_{\mathbf p}}
\int_\Sigma
\omega(x,p)\,W(\Pi^2)\,S^\mu(x,p),
\end{aligned}
\end{equation}
where $\omega(x,p)$ is defined in Eq.~\eqref{eq:omega_def}. Consequently,
the factor $E_{\mathbf p}^{-1}$, which cancels at fixed momentum
in Eq.~\eqref{eq:Pmu_as_weighted_avg}, must be retained when the spectra are integrated
over a finite momentum-space acceptance. Eq.~\eqref{eq:Pmu_accept_expanded} therefore
takes the form
\begin{equation}
P^\mu_{\rm GUP}[\mathcal A]
=
2\,
\frac{
\displaystyle
\int_{\mathcal A}\frac{d^3p}{E_{\mathbf p}}
\int_\Sigma
\omega(x,p)\,W(\Pi^2)\,S^\mu(x,p)
}{
\displaystyle
\int_{\mathcal A}\frac{d^3p}{E_{\mathbf p}}
\int_\Sigma
\omega(x,p)\,W(\Pi^2)
}.
\end{equation}
It is therefore convenient to define the undeformed freeze-out
average over $(x,p)\in\Sigma\times\mathcal A$ by
\begin{equation}
\left\langle X\right\rangle_{\rm fo,\mathcal A}
\equiv
\frac{
\displaystyle
\int_{\mathcal A}\frac{d^3p}{E_{\mathbf p}}
\int_\Sigma
\omega(x,p)\,X(x,p)
}{
\displaystyle
\int_{\mathcal A}\frac{d^3p}{E_{\mathbf p}}
\int_\Sigma
\omega(x,p)
\label{eq.6.17}
}.
\end{equation}
The momentum element $d^3p/E_p$, together with the factor
$(2\pi)^{-3}$ included in $\omega(x,p)$, reproduces the
positive-energy Lorentz-invariant on-shell phase-space element defined in
Eq.~\eqref{eq.3.1}. The common normalization cancels from the normalized
average in Eq.~\eqref{eq.6.17}. With the acceptance integration defined
consistently in Eq.~\eqref{eq.6.17}, the GUP-modified polarization can be
written compactly as}
\begin{equation}
P^\mu_{\rm GUP}[{\cal A}]
=
2\,
\frac{\left\langle W(\Pi^2)\,S^\mu\right\rangle_{\rm fo,{\cal A}}}
{\left\langle W(\Pi^2)\right\rangle_{\rm fo,{\cal A}}}.
\label{eq:Pmu_accept_weighted}
\end{equation}
{ Expanding the scalar GUP factor consistently to first order,
\(
W(\Pi^2)
=
1-\kappa\beta_G\Pi^2
+\mathcal{O}(\beta_G^2\Pi^4),
\)
one obtains
\begin{align}
P^\mu_{\rm GUP}[\cal A]
&=
2\,
\frac{
\left\langle
(1-\kappa\beta_G\Pi^2)S^\mu
\right\rangle_{\rm fo,\cal A}
}{
\left\langle
1-\kappa\beta_G\Pi^2
\right\rangle_{\rm fo,\cal A}
}
\nonumber\\
&=
2\Big[
\left\langle S^\mu\right\rangle_{\rm fo,\cal A}
-\kappa\beta_G
\Big(
\left\langle\Pi^2S^\mu\right\rangle_{\rm fo,\cal A}
-\nonumber\\
&\hspace{4cm}\left\langle\Pi^2\right\rangle_{\rm fo,\cal A}
\left\langle S^\mu\right\rangle_{\rm fo,\cal A}
\Big)
\Big]
\nonumber\\
&=
P^\mu_{\rm std}[\cal A]
-
2\kappa\beta_G\,
{\rm Cov}_{\rm fo,\cal A}
\!\left(\Pi^2,S^\mu\right)
+\nonumber\\
&\hspace{4.5cm}
\mathcal{O}
\!\left(
\beta_G^2\partial,
\partial^2,
\beta_G\partial^2
\right).
\label{eq.6.20}
\end{align}
where
\(
P^\mu_{\rm std}[\cal A]
\equiv
2\left\langle S^\mu\right\rangle_{\rm fo,\cal A},
\)
and the acceptance-matched covariance is defined by
\begin{equation}
{\rm Cov}_{\rm fo,\cal A}
\!\left(\Pi^2,S^\mu\right)
\equiv
\left\langle\Pi^2S^\mu\right\rangle_{\rm fo,\cal A}
-
\left\langle\Pi^2\right\rangle_{\rm fo,\cal A}
\left\langle S^\mu\right\rangle_{\rm fo,\cal A}.
\label{eq:cov_foA_def}
\end{equation}
}
{Thus, the invariant momentum factor required in the acceptance integration
changes the precise definition of the freeze-out average but leaves the
covariance structure of the leading GUP correction unchanged. For the fully momentum-integrated global polarization, $\cal A$ is taken to be the full detector acceptance, $\cal A\rightarrow \cal A_{\mathrm{all}}$, and Eq.~\eqref{eq.6.20} remains valid. Eq.~\eqref{eq.6.20} is the central observable result of the present analysis. The common scalar phase-space factor cancels from the local spin ratio but need not cancel after freeze-out integration because $\Pi^2(x,p)$ and $S^\mu(x,p)$ can be correlated over the accepted freeze-out domain. The leading correction vanishes when this covariance vanishes. The pseudogauge and spin-observable scope of this result is precisely
that specified in Eq.~\eqref{eq.5.15}.
}
Experiments typically quote the polarization in the particle rest frame. If $P^\mu(p)$ is known
in the lab frame, the spatial rest-frame polarization can be obtained via
\begin{equation}
\bm{P}^{\ast}(p)
=
\bm{P}(p)
-\frac{\bm{p}}{E_p+m}\,P^0(p),
\label{eq:Prest_from_P}
\end{equation}
using $P\!\cdot\!p=0$ at leading order in gradients \cite{Becattini:2016gvu}.
The global polarization reported by STAR corresponds to the component of $\bm{P}^{\ast}$
along the system angular momentum direction in non-central collisions \cite{STAR:2017ckg}.

\section{Phenomenological sensitivity and theoretical uncertainty}
\label{sec:bounds}

{

Eq.~\eqref{eq.6.20} shows that a quantitative
constraint on $\beta_G$ requires the acceptance-matched covariance
between $\Pi^2$ and the spin component appearing in the observable.
To estimate the characteristic magnitude of this covariance without
identifying a simplified thermal estimate with a full hydrodynamic
calculation, consider a $\Lambda$ hyperon in the LRF,
with
$f_{\rm MB}(p)\propto \exp(-E_{\bm p}/T_f)$,
$E_{\bm p}=\sqrt{\bm p^2+m^2}$, and
$S_0(p)\simeq\chi E_{\bm p}$.
Here, we define $\Delta P\equiv P_{\rm GUP}-P_{\rm std}$, the relative
correction becomes
\begin{multline}
\left.
\frac{\Delta P}{P_{\rm std}}
\right|_{\rm MB}
=
-\kappa\beta_G\,\mathcal R(m,T_f)
+\mathcal O(\beta_G^2,\beta_G\partial),
\\
\mathcal R(m,T_f)
\equiv
\frac{
\operatorname{Cov}_{T_f}(\bm p^2,E_{\bm p})
}{
\langle E_{\bm p}\rangle_{T_f}
}.
\label{eq:relative_MB_compact}
\end{multline}
Here
\begin{multline}
\langle X\rangle_{T_f}
=
\frac{
\displaystyle
\int d^3\bm p\,
X(\bm p)e^{-E_{\bm p}/T_f}
}{
\displaystyle
\int d^3\bm p\,
e^{-E_{\bm p}/T_f}
},
\\
\operatorname{Cov}_{T_f}(A,B)
=
\langle AB\rangle_{T_f}
-
\langle A\rangle_{T_f}
\langle B\rangle_{T_f}.
\label{eq:thermal_average_compact}
\end{multline}
With $z\equiv m/T_f$, the standard Maxwell-Boltzmann integrals give
\cite{deGroot1980}
\begin{equation}
\mathcal R(m,T_f)
=
\frac{
3m^2T_f
\left[
1+\dfrac{2}{z}\dfrac{K_3(z)}{K_2(z)}
-\dfrac{K_1(z)K_3(z)}{K_2^2(z)}
\right]
}{
3T_f+m\,\dfrac{K_1(z)}{K_2(z)}
}.
\label{eq:R_MB_compact}
\end{equation}
The asymptotic limits are
\begin{equation}
\mathcal R(m,T_f)
\longrightarrow
\begin{cases}
3T_f^2, & m/T_f\gg1,\\[2mm]
8T_f^2, & m/T_f\ll1.
\end{cases}
\label{eq:R_controlled_limits}
\end{equation}
For $m_\Lambda=1.115683~\mathrm{GeV}$,
Eq.~\eqref{eq:R_MB_compact} gives
\begin{multline}
\mathcal R(m_\Lambda,T_f)
=
0.0389,\ 0.0583,\ 0.1115~\mathrm{GeV}^2
\quad\text{for}\\
T_f=0.10,\ 0.12,\ 0.16~\mathrm{GeV},
\label{eq:R_lambda_values}
\end{multline}
respectively. Across this representative freeze-out interval
\cite{Andronic_Nature2018,Becattini:2016gvu},
$\mathcal R$ changes by a factor of $2.87$, while
$1/(\kappa\mathcal R)$ changes by the inverse factor.

Replacing the Maxwell-Boltzmann distribution by the Fermi-Dirac
distribution at $\mu_B\simeq0$ changes $\mathcal R$ by less than
$10^{-4}$ in relative terms over the same temperature interval.
This small difference follows from the dilute occupation of a
$\Lambda$ mode for $m_\Lambda/T_f\simeq7$-$11$. The statistical
choice is therefore numerically negligible here compared with the
dependence on $T_f$ and on the spacetime correlations of the fluid
fields. 
To illustrate the role of transverse flow, consider a midrapidity
profile
\begin{equation}
E_*(p_T,\Delta\phi)
=
\gamma_T
\left(
m_T-\bar\beta_Tp_T\cos\Delta\phi
\right),
\label{eq:flow_check_energy}
\end{equation}
where $\gamma_T=(1-\bar\beta_T^2)^{-1/2},$ and
\begin{equation}
d\mu_{\mathcal A}
\propto
p_Tm_T\,e^{-E_*/T_f}\,
dp_T\,d\Delta\phi.
\label{eq:flow_measure}
\end{equation}
For this illustration, the LRF momentum invariant is
$\Pi^2=E_*^2-m^2$, while the corresponding spin factor is taken to be
proportional to $E_*$. At $T_f=0.12~\mathrm{GeV}$ and
$\bar\beta_T=0.60$, the analogue of $\mathcal R$ changes from
$0.0437~\mathrm{GeV}^2$ in the no-flow approximation to
$0.0486~\mathrm{GeV}^2$ for
$0.5<p_T<5~\mathrm{GeV}/c$, a difference of about $11\%$.
The difference can be substantially larger in narrower $p_T$
intervals. This comparison is illustrative and does not define a
complete theoretical uncertainty interval. Event-by-event viscous
hydrodynamics may further modify the correlations among
$u^\mu(x)$, $T(x)$, $\varpi_{\mu\nu}(x)$, the freeze-out
hypersurface, and resonance feed-down.  
{The relative importance of the freeze-out temperature, quantum-statistical
correction, and the illustrative transverse-flow effect is summarized in
Fig.~\ref{fig:R_Tf_sensitivity}.
\begin{figure}[t]
\centering
\includegraphics[width=0.98\columnwidth,height=0.82\columnwidth]{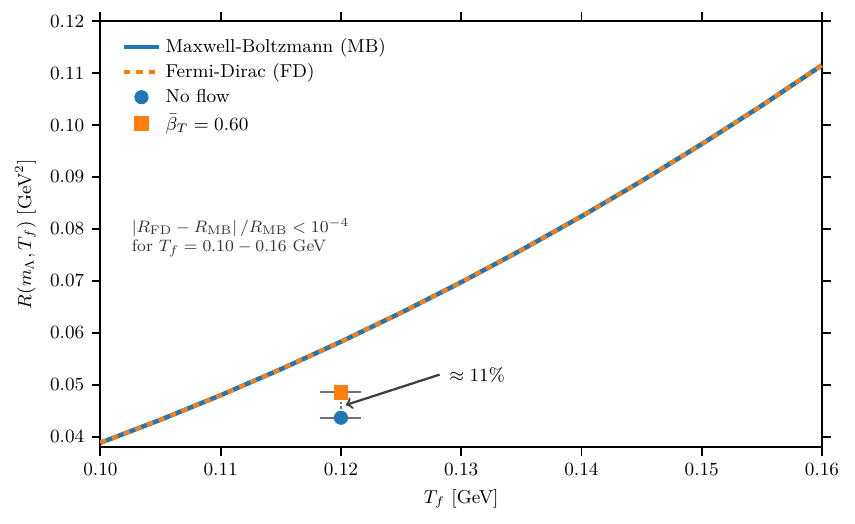}

\caption{\justifying
Dependence of the thermal mapping quantity $R(m_\Lambda,T_f)$ on
the freeze-out temperature. The solid blue and dashed orange curves
show the full-thermal Maxwell-Boltzmann and Fermi-Dirac results,
respectively, their relative difference remains below $10^{-4}$ over
$T_f=0.10$-$0.16$ GeV. The circle and square at $T_f=0.12$ GeV show
the illustrative acceptance-restricted results for $0.5<p_T<5$ GeV/$c$
without transverse flow and with $\bar{\beta}_T=0.60$, respectively,
corresponding to an approximately $11\%$ change. Within this benchmark,
the freeze-out-temperature dependence and the illustrative flow effect
are substantially larger than the quantum-statistical correction.
The flow comparison is not intended as a complete hydrodynamic
uncertainty estimate.
}

\label{fig:R_Tf_sensitivity}

\end{figure}
}
These correlations are
therefore a major source of theoretical uncertainty when
Eq.~\eqref{eq.6.20} is converted into a numerical
estimate. 
For a restricted momentum domain $\mathcal A$, the no-flow thermal
average is
\begin{equation}
\langle X\rangle_{T_f,\mathcal A}
=
\frac{
\displaystyle
\int_{\mathcal A}
dy\,dp_T\,d\phi\;
p_TE_{\bm p}\,
X(p_T,y,\phi)e^{-E_{\bm p}/T_f}
}{
\displaystyle
\int_{\mathcal A}
dy\,dp_T\,d\phi\;
p_TE_{\bm p}\,
e^{-E_{\bm p}/T_f}
}.
\label{eq:acceptance_average_corrected}
\end{equation}
Here \( E_{\bm p}=m_T\cosh y,
\quad m_T=\sqrt{m^2+p_T^2}.
\)
The factor $E_{\bm p}$ in
Eq.~\eqref{eq:acceptance_average_corrected} follows from the
Jacobian
\begin{equation}
d^3\bm p
=
p_TE_{\bm p}\,
dp_T\,dy\,d\phi.
\end{equation}
The corresponding acceptance-dependent quantity is given in Eq. 
\begin{equation}
\mathcal R_{\mathcal A}(m,T_f)
=
\frac{
\operatorname{Cov}_{T_f,\mathcal A}
(\bm p^2,E_{\bm p})
}{
\langle E_{\bm p}\rangle_{T_f,\mathcal A}
}.
\label{eq:R_acceptance_compact}
\end{equation}
Eq.~\eqref{eq:relative_MB_compact} then remains valid after the
replacement
$\mathcal R\rightarrow\mathcal R_{\mathcal A}$.
In a general hydrodynamic freeze-out calculation, however, the
complete covariance in Eq.~\eqref{eq:cov_foA_def} must be
evaluated over both the hypersurface and the accepted momentum
domain. 
Let $\delta P_{\rm exp}[\mathcal A]$ denote the total experimental
uncertainty for a fixed collision system, centrality interval, and
acceptance. Within the Maxwell-Boltzmann approximation, the linear
sensitivity to $\beta_G$ is given by
\begin{equation}
\left|\beta_G\right|_{\rm sens}
\equiv
\frac{
\delta P_{\rm exp}[\mathcal A]
}{
\kappa\mathcal R_{\mathcal A}(m,T_f)
\left|P_{\rm std}[\mathcal A]\right|
}.
\label{eq:formal_linear_sensitivity}
\end{equation}
The corrected ALICE average for global hyperon polarization
in the $15$-$50\%$ centrality interval is
\cite{ALICE:2019onw}
\begin{multline}
\langle P_H\rangle
=
-0.01
\pm0.05\,\mathrm{(stat.)}
\pm0.03\,\mathrm{(syst.)}\ \%,
\\
\delta P_{\rm tot}
=
\sqrt{0.05^2+0.03^2}\ \%
\simeq0.058\ \%.
\label{eq:ALICE_corrected}
\end{multline}
If the illustrative reference magnitude
$\left|P_{\rm std}\right|=0.01\%$ is adopted together with the value
$\kappa=5$ used in this work,
Eqs.~\eqref{eq:R_lambda_values} and
\eqref{eq:formal_linear_sensitivity} give
\begin{equation}
\left|\beta_G\right|_{\rm sens}^{(1\sigma)}
=
\frac{0.058/0.01}
{5\,\mathcal R(m_\Lambda,T_f)}
\simeq
10\text{--}30~\mathrm{GeV}^{-2}.
\label{eq:beta_formal_sensitivity}
\end{equation}
Multiplying only the experimental uncertainty by $1.96$ gives
\begin{equation}
\left|\beta_G\right|_{\rm sens}^{(1.96\sigma)}
\simeq
20\text{--}59~\mathrm{GeV}^{-2}.
\label{eq:beta_formal_sensitivity_196}
\end{equation}
Eqs.~\eqref{eq:beta_formal_sensitivity} and
\eqref{eq:beta_formal_sensitivity_196} should be interpreted only as
sensitivity estimates, not as exclusion limits. Both the reference
polarization and the relevant covariance depend on the freeze-out
description and, more importantly, the inferred values of $\beta_G$
lie far outside the domain in which the linear expansion is valid. 
The first-order expansion requires
\begin{equation}
\epsilon_G(x,p)
\equiv
\kappa\left|\beta_G\right|\Pi^2(x,p)
\ll1
\label{eq:perturbative_control}
\end{equation}
throughout the momentum region that contributes appreciably to the
observable. In the same thermal approximation,
\begin{equation}
\langle\Pi^2\rangle_{T_f}
=
3m_\Lambda T_f
\frac{
K_3(m_\Lambda/T_f)
}{
K_2(m_\Lambda/T_f)
}
=
0.414,\ 0.517,\ 0.745~\mathrm{GeV}^2
\label{eq:Pi2_thermal_values}
\end{equation}
for $T_f=0.10$, $0.12$, and $0.16~\mathrm{GeV}$, respectively.
For $\kappa=5$, applying
Eq.~\eqref{eq:perturbative_control} even at the mean value gives
\( \left|\beta_G\right| \ll 0.483,\ 0.386,\ 0.268~\mathrm{GeV}^{-2}\) 
at the same three temperatures, with a stronger restriction in the
high-momentum tail. In contrast, the sensitivity interval
$\left|\beta_G\right|=10$-$30~\mathrm{GeV}^{-2}$ implies
\(
\kappa\left|\beta_G\right|
\langle\Pi^2\rangle
\simeq
21\text{-}112,
\)
which is incompatible with
Eq.~\eqref{eq:perturbative_control}.
The ALICE precision represented by
Eq.~\eqref{eq:ALICE_corrected} therefore does not exclude any value
of $\beta_G$ within the reliable first-order domain. For the KMM
minimal-length interpretation, $\beta_G\geq0$, the absolute-value
notation in the sensitivity expressions is used only to display the
magnitude of a generic effective linear coefficient. 
The usefulness of a polarization observable is determined by the
magnitude of
\(
\left|
\operatorname{Cov}_{{\rm fo},\mathcal A}
(\Pi^2,S_O)
\right|
\)
relative to the combined experimental and hydrodynamic uncertainty.
Momentum-, rapidity-, centrality-, and azimuthally resolved
polarization measurements are therefore more informative than a
single fully integrated value, since integration can cancel
contributions of opposite sign. Differential global polarization
and the longitudinal coefficients $P_{z,s2}$ and $P_{z,s3}$ provide
direct tests of the covariance structure predicted here.

For the global component, existing RHIC measurements offer a more
favorable comparison because the measured $\Lambda$ polarization is
nonzero and larger than the nearly vanishing global polarization
reported at LHC energies
\cite{STAR:2017ckg,ALICE:2019onw}.
ALICE, on the other hand, provides particularly useful information
through the $p_T$, rapidity, centrality, and azimuthal dependence of
longitudinal polarization. The measurement of $P_{z,s2}$ at
$\sqrt{s_{\rm NN}}=5.02~\mathrm{TeV}$ and the measurements of both
$P_{z,s2}$ and $P_{z,s3}$ at
$\sqrt{s_{\rm NN}}=5.36~\mathrm{TeV}$ provide complementary harmonic
information
\cite{ALICE2022PRLz,ALICE:2026eje}.
Thus, RHIC is more favorable for a global-amplitude comparison,
whereas the larger ALICE data samples are especially valuable for
differential harmonic comparisons. 
Vector-meson spin alignment, conventionally described by the
spin-density matrix element $\rho_{00}$, lies outside the spin-$1/2$
Pauli-Lubanski/axial-Wigner framework used here. Extending the
present analysis to spin alignment would require a separate spin-1
density-matrix formulation. Accordingly, no constraint from
$\rho_{00}$ is claimed in this work.

}

\section{Conceptual Consistency}
\label{sec8}

This section states explicitly the assumptions underlying the preceding
derivations and identifies the additional elements that would be needed
for a fully Lorentz-covariant quantum-field theoretic formulation of the
GUP. The scalar GUP framework and the fixed microscopic operator content
adopted in the preceding sections are summarized in
Secs.~\ref{sec3}-\ref{sec5}. Here we examine only
the relativistic-frame and pseudogauge scope of those assumptions.
The isotropic GUP algebra introduced in Eq.~\eqref{eq:GUP_general} is not manifestly
Lorentz covariant because it singles out spatial components.
A relativistic implementation therefore requires specifying the frame
in which that algebra is imposed.  
Consequently, the present analysis is restricted to a fixed local rest
frame, and the resulting polarization effects should be understood within
an effective GUP description of the local phase-space structure. Fully Lorentz-covariant minimal-length completions (for example, Snyder-type or DSR-inspired deformed algebras with modified generators and transformation laws) may modify the resulting polarization corrections, potentially already at $\mathcal{O}(\beta_G)$, including the cancellation of the pointwise mean-spin kernel derived in Sec.~\ref{sec5}.  
In a relativistic fluid, a timelike unit vector field is already present, namely the local flow velocity $u^\mu(x)$, and it is natural to identify the GUP frame with the LRF determined by $u^\mu$ (equivalently, the $\beta$ frame) \cite{Becattini:2014yxa}. 
An orthonormal tetrad $e_a^{\ \mu}(x)$ with
\begin{equation}
e_0^{\ \mu}(x)=u^\mu(x),
\quad
e_a^{\ \mu} e_{b\mu}=\eta_{ab},
\label{eq:tetrad_def}
\end{equation}
where $\eta_{ab}=\mathrm{diag}(1,-1,-1,-1),$ and the local components $p^{\hat{i}} \equiv e^{\hat{i}}_{\ \mu} p^\mu$ for $\hat{i}=1,2,3$. 
Using the definition of $\Pi^2$ in Eq.~\eqref{eq:Pi2_cov}, its LRF
tetrad representation can equivalently be written as
\begin{equation}
\Pi^2
=
\delta_{\hat i\hat j}\,
p^{\hat i}p^{\hat j}
=
-\Delta_{\mu\nu}p^\mu p^\nu,
\quad
\Delta_{\mu\nu}
\equiv
g_{\mu\nu}-u_\mu u_\nu .
\end{equation}
With this choice, the GUP-dependent phase-space structure can be expressed
in a manifestly covariant form,
\begin{equation}
dP_{\rm GUP}=dP\,\mathcal{W}(\Pi^2),
\label{eq:covariant-gup-measure}
\end{equation}
where $dP$ is the positive-energy invariant on-shell measure defined in
Eq.~\eqref{eq.3.1}. Since both $dP$ and $\Pi^2$ are Lorentz scalars,
the expression above is covariant once the hydrodynamic four-velocity $u^\mu$
specifying the LRF is transformed together with $p^\mu$.
Although Eq. \eqref{eq:covariant-gup-measure} is covariant in the sense above, it still uses $u^\mu$ as a physical timelike direction and thus corresponds to an effective theory with a built-in LRF. This is appropriate for hydrodynamic applications, but it is distinct from a fundamental Lorentz covariant deformation of spacetime. A fully Lorentz covariant algebra exists, for instance Snyder type noncommutative spacetime \cite{Snyder1947} and Lorentz covariant deformed algebras with minimal length \cite{QuesneTkachuk2006}, which schematically take the form
\begin{equation}
[x_\mu,x_\nu] \sim i\,\ell^2\,M_{\mu\nu},
\quad
[x_\mu,p_\nu] \sim i\left(g_{\mu\nu}+\ell^2 p_\mu p_\nu\right),
\label{eq:Snyder_like}
\end{equation}
with $M_{\mu\nu}$ the Lorentz generators. Such covariant completions generally entail additional operator structure and modified transformation laws, and they are not assumed here. 
In dissipative hydrodynamics, the definition of $u^\mu$ (Landau, Eckart, or $\beta$ frame) differs by gradient corrections. If $u^\mu \to u^\mu + \delta u^\mu$ with $\delta u^\mu=\mathcal{O}(\partial)$, then $\Pi^2=(p\!\cdot\!u)^2-m^2$ changes by $\delta\Pi^2=\mathcal{O}(\partial)$, hence the induced change in $\mathcal{W}(\Pi^2)$ is $\mathcal{O}(\beta_G\partial)$. Since the polarization kernel itself is already $\mathcal{O}(\partial)$, the resulting difference in the final polarization is $\mathcal{O}(\beta_G\partial^2)$, which lies beyond the truncation adopted here.
{ The local separation of total angular momentum into orbital and spin
parts is not unique. Under a pseudogauge transformation generated by
$\Phi^{\lambda,\mu\nu}=-\Phi^{\lambda,\nu\mu}$,
\begin{align}
T_{\mathfrak p'}^{\mu\nu}
&=T_{\mathfrak p}^{\mu\nu}
+\frac{1}{2}\partial_\lambda
\left(
\Phi^{\lambda,\mu\nu}
+\Phi^{\mu,\nu\lambda}
+\Phi^{\nu,\mu\lambda}
\right),\\
S_{\mathfrak p'}^{\lambda,\mu\nu}
&=S_{\mathfrak p}^{\lambda,\mu\nu}
-\Phi^{\lambda,\mu\nu}.
\end{align}
As noted after Eq.~\eqref{eq.4.15}, the integrated charges agree when the
corresponding boundary terms vanish. This global equality does not make
the local energy-momentum tensors, spin tensors, or local spin
observables identical.

A local-equilibrium operator in a general pseudogauge can contain both
energy-momentum and spin-tensor constraints,
\begin{multline}
\widehat\rho_{\mathrm{LE}}^{(\mathfrak p)}
=\frac{1}{Z_{\mathfrak p}}
\exp\Big[
-\int_\Sigma d\Sigma_\mu
\Big(
\widehat T_{\mathfrak p}^{\mu\nu}\beta_\nu
-\frac{1}{2}\Omega_{\alpha\beta}
\widehat S_{\mathfrak p}^{\mu,\alpha\beta}
-\\\sum_a\xi_a\widehat j_a^\mu
\Big)
\Big],
\end{multline}
where the spin potential and matching conditions form part of the
chosen local theory. Local-equilibrium polarization can consequently
differ between pseudogauges. 
Recent results make this distinction physically relevant. For a free
Dirac field, among the nontrivial integrated spin operators studied in
the canonical, de Groot-van Leeuwen-van Weert, and
Hilgevoord-Wouthuysen pseudogauges, only the canonical spin operator
satisfies the $SO(3)$ angular-momentum algebra~\cite{Dey:2023hft}. Smeared equal-time
commutators of local energy-momentum components are also pseudogauge
dependent at finite coarse-graining width and approach
pseudogauge-independent behavior only in the appropriate large-scale
limit~\cite{Dey:2024crk}. Pseudogauge transformations therefore cannot generally be
treated as physically empty redefinitions of local quantum observables.

The present calculation fixes $\mathfrak p=\mathrm{BR}$, uses the
symmetric Belinfante-Rosenfeld tensor in
$\widehat\rho_{\mathrm{LE}}$, and defines the detected one-particle spin
through the axial Wigner component, equivalently the one-particle
Pauli-Lubanski operator. In this convention the BR spin tensor vanishes
as an independent hydrodynamic constraint, but the detected particle
continues to carry the standard Pauli-Lubanski spin. 
The cancellation established in Sec.~\ref{sec5} follows only if the same
spin-independent scalar factor multiplies the spin numerator and scalar
denominator within this fixed convention. It neither implies equality
of spin observables in two pseudogauges nor proves that the required
factorization survives a change to the canonical spin tensor, a nonzero
spin potential, or different matching conditions. Repeating the
construction in another pseudogauge is a distinct calculation and lies
outside the scope of this work.
}

\begin{figure}[htb]
\centering
\includegraphics[width=0.98\columnwidth,height=0.82\columnwidth]{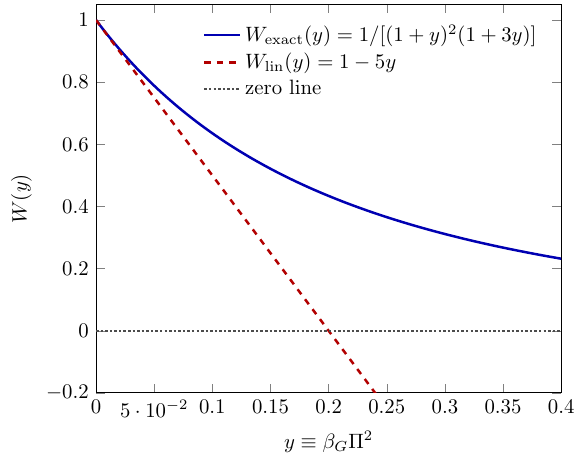}
\caption{\justifying Comparison of the exact isotropic GUP weight for the Jacobi-consistent realization, $\mathcal{W}_{\rm exact}(y)=1/[(1+y)^2(1+3y)]$ with $y\equiv \beta_G\Pi^2$, and its linear truncation $\mathcal{W}_{\rm lin}(y)=1-5y$ used in the $\mathcal{O}(\beta_G)$ expansion. The dotted horizontal line marks $\mathcal{W}=0$. The exact weight remains positive for $\beta_G>0$, while the linear approximation crosses zero at $y=0.2$, illustrating that the linearized treatment is meaningful only when $y=\beta_G\Pi^2\ll 1$ over the dominant momentum support. This figure therefore visualizes the positivity and controlled-expansion requirements that underlie the perturbative use of
$\mathcal{W}(\Pi^2)$ in the freeze-out integrals.}
\label{fig:Wexact_vs_linear}
\end{figure}

Pseudo gauge transformations play a central role in hydrodynamics with spin degrees of freedom. Their implications for local thermodynamics and higher-order transport have been analyzed in detail, including how pseudo gauge changes can be reinterpreted as field redefinitions at second order in gradients \cite{Li:2020eon,Fukushima:2020ucl,Hongo:2021ona}. Since the present paper truncates at first order in gradients, these higher-order subtleties are not resolved here, and the pseudo gauge must be fixed as part of the definition of the effective theory.  
The standard mass shell and dispersion relation are already fixed by
Eq.~\eqref{eq.3.1} and form part of the scope stated in Eq.~\eqref{eq.5.15}.
Accordingly, the present scalar GUP framework does not modify
single-particle kinetic propagation. Potential microcausality issues may
arise in fully quantized nonlocal formulations of modified algebras, but
their assessment requires a specified underlying quantum field theory and
lies beyond the effective description considered here
\cite{Hossenfelder:2012jw}. The corresponding modification of the density
of states is represented by $\mathcal{W}(\Pi^2)$.
For the exact rational factor associated with the Jacobi-consistent phase-space structure in Eq.~\eqref{eq:measure_beta2beta}, $W(\Pi^2)>0$ for $\beta_G>0$ and $\Pi^2\geq0$, whereas its first-order expansion is given by Eq.~\eqref{eq:W_general}.  Fig.~\ref{fig:Wexact_vs_linear} compares the exact and linear forms.
One must impose the stronger condition
\(
\kappa\,\beta_G\,\Pi^2\ll 1\: \text{on the dominant support of the freeze-out integrals},
\)
so that $\mathcal{W}$ remains positive and the truncation is controlled. A characteristic freeze-out momentum is specified through $\Lambda_{\rm fo}^2\sim\langle\Pi^2\rangle_{\rm fo}$, so that the dimensionless expansion parameter is \(\epsilon_\beta\equiv\kappa\,\beta_G\,\Lambda_{\rm fo}^2\ll1 .\) The polarization terms retained in the present analysis are therefore of relative order $\epsilon_\beta$ with respect to the standard result.
Because the present GUP framework modifies the phase-space structure
without specifying a corresponding Hamiltonian or interaction vertices,
no conclusion is drawn regarding the unitarity of a GUP-modified
$S$ matrix. At the effective level relevant to freeze-out observables, we assume that
the GUP modifies the phase-space structure universally. A microscopic
formulation that preserves unitarity would also need to specify the
operator representation and the associated symmetry generators, as in
Snyder-like or doubly special relativity formulations
\cite{Snyder1947,AmelinoCamelia2002,MagueijoSmolin2002,QuesneTkachuk2006}.
The applicable range of the present analysis is therefore restricted by
the assumptions summarized in Eq.~\eqref{eq.5.15} and by the perturbative
condition in Eq.~\eqref{eq:perturbative_control}. Outside this range,
particularly for fully covariant formulations of the GUP or in
strong-gradient regions where shear contributions become comparable to
those from thermal vorticity, the microscopic operator structure must be
specified more completely.

{
\section{Conclusion}
\label{sec9}
We have studied an isotropic GUP introduced as a scalar modification of
the LRF phase-space structure within a fixed Belinfante-Rosenfeld
Zubarev local-equilibrium formalism. With the standard mass shell and the
Pauli-Lubanski/axial-Wigner spin observable left unchanged, the same
factor $\mathcal{W}(\Pi^2)$ multiplies the scalar and axial phase-space densities.
It therefore cancels from the local mean-spin ratio at fixed $(x,p)$
through $\mathcal{O}(\beta_G\partial)$. This cancellation holds only
under the assumptions adopted here and should not be interpreted as a
statement of pseudogauge invariance. A different pseudogauge, spin
operator, spin potential, dispersion relation, momentum-space
anisotropy, or perturbative order can modify the local spin structure
and remove the cancellation. 
Although the scalar factor cancels locally, it can affect the observed
polarization after Cooper-Frye integration. The leading
momentum-differential and acceptance-averaged contributions are expressed
through covariances between $\Pi^2$ and the thermal-vorticity spin kernel
over the freeze-out hypersurface and the accepted momentum region. For
$\Lambda$ hyperons in the Maxwell-Boltzmann approximation, varying
$T_f$ from $0.10$ to $0.16~\mathrm{GeV}$ changes the relevant thermal
moment by a factor of $2.87$. The Fermi-Dirac correction is negligible
over this interval, whereas correlations associated with the
hydrodynamic evolution, freeze-out hypersurface, and resonance feed-down
remain the dominant theoretical uncertainties. The present ALICE
precision corresponds formally to
$|\beta_G|\sim10$-$30~\mathrm{GeV}^{-2}$, however, this interval does
not satisfy the perturbative condition
$\kappa|\beta_G|\Pi^2\ll1$. Present data therefore provide only a formal
estimate of sensitivity and do not support a controlled exclusion limit. 
The most informative future observables are momentum- and
acceptance-resolved polarization measurements, together with rapidity,
centrality, and azimuthal-harmonic dependences, because the predicted GUP
contribution is governed by freeze-out covariances rather than by an
additional local tensor structure. RHIC currently provides stronger
sensitivity through the global polarization amplitude, whereas
high-statistics ALICE measurements of longitudinal polarization
harmonics offer a complementary test of the differential momentum and
azimuthal dependence. A quantitatively controlled constraint on
$\beta_G$ will require event-by-event hydrodynamic calculations using
the same experimental acceptance as the polarization data, together
with sufficient experimental precision to ensure that the inferred
dimensionless expansion parameter remains well below unity.
}
\section*{Acknowledgments}
The author SAM is thankful to Mir Faizal of Durham University for fruitful discussions and understanding of many aspects of the Quantum Gravity.

\section*{Conflict of Interest}
The authors declare that there are no conflicts of interest regarding this work.

\section*{Data Availability Statement}
No external datasets were produced in this study. All numerical inputs
and theoretical results required to reproduce the analysis are provided
in the manuscript and the cited references.

\bibliographystyle{apsrev4-1}
\bibliography{refs_rf_gup-1}

\end{document}